\documentclass[12pt,preprint,letterpaper]{aastex}

\newcommand{\km}{\,{\textnormal{km}}}
\newcommand{\second}{\,{\textnormal{s}}}
\newcommand{\gram}{\,{\textnormal{g}}}
\newcommand{\cm}{\,{\textnormal{cm}}}
\newcommand{\kms}{{\km\second^{-1}}}
\newcommand{\solarmass}{{\,{\textnormal{M}}_\sun}}
\newcommand{\yr}{{\,\textnormal{yr}}}
\newcommand{\solarmassyear}{{\solarmass\yr^{-1}}}
\newcommand{\AU}{{\,\textnormal{AU}}}

\newcommand{\gauss}{\,{\textnormal{G}}}
\newcommand{\massden}{{\gram\cm^{-3}}}
\newcommand{\numden}{{\cm^{-3}}}
\def\lsim{\raise0.3ex\hbox{$<$}\kern-0.75em{\lower0.65ex\hbox{$\sim$}}}
\def\gsim{\raise0.3ex\hbox{$>$}\kern-0.75em{\lower0.65ex\hbox{$\sim$}}}

\begin{document}

\shorttitle{MAGNETOCENTRIFUGAL WINDS I - STEADY MASS LOADING}
\shortauthors{ANDERSON ET AL.}

\title{The Structure of Magnetocentrifugal Winds: I. Steady Mass Loading}

\author{Jeffrey M. Anderson, Zhi-Yun Li}
\affil{Department of Astronomy, University of Virginia, P. O. Box 3818,
Charlottesville, VA 22903}
\email{jma2u@virginia.edu, zl4h@virginia.edu}
\author{Ruben Krasnopolsky}
\affil{Center for Theoretical Astrophysics, University of Illinois at
Urbana-Champaign, Loomis Laboratory, 1110 West Green Street, 
Urbana, IL 61801}
\email{ruben@astro.uiuc.edu}
\and
\author{Roger D. Blandford}
\affil{SLAC, M/S 75, 2575 Sandhill Rd, Menlo Park, CA 94025}
\email{rdb3@stanford.edu}

\begin{abstract}
We present the results of a series of time-dependent numerical simulations
of cold, 
magnetocentrifugally launched winds from accretion disks.  The goal of
this study is to determine how the mass loading from the disk affects the
structure and dynamics of the wind for a given distribution of magnetic
field.  Our simulations span four and half decades of
mass loading; in the context of a disk with a launching region from $0.1\AU$
to $1.0\AU$ around a $1\solarmass$ star and a field strength of about 
$20\gauss$
at the inner disk edge, this amounts to mass loss rates of 
$1\times 10^{-9}$ -- $3\times 10^{-5}\solarmassyear$ from each side of the 
disk.  We find that, as expected intuitively,  
the degree of collimation of the wind increases with mass loading; however
even the ``lightest'' wind simulated is significantly collimated compared
with the force-free magnetic configuration of the same magnetic flux
distribution at the launching surface, which becomes radial at large
distances.  The
implication is that for flows from young stellar objects a radial field
approximation is inappropriate.  Surprisingly, the terminal velocity of the
wind and the magnetic lever arm are still well-described by the analytical
solutions for a radial field geometry.  We also find that the isodensity
contours and Alfv\'en surface are approximately self-similar in mass 
loading. The wind becomes
unsteady above some critical mass loading rate.  The exact value of the 
critical rate depends on the (small) velocity with which we inject the
material into the wind. For a small enough injection speed, we are 
able to obtain the first examples of a class of heavily-loaded 
magnetocentrifugal winds with magnetic fields completely dominated 
by the toroidal component all the way to the launching surface. 
The stability of such toroidally dominated winds in 3D will be
the subject of a future investigation. 
\end{abstract}

\keywords{ISM: jets and outflows  --- MHD --- stars: formation --- stars: 
mass loss --- stars: pre-main-sequence}

\section{INTRODUCTION}

\subsection{Magnetocentrifugal Wind Solutions}

Astrophysical jets are seen in a variety of systems ranging from young
stellar objects to active galactic nuclei.  All these systems
with jets are also observed to have accretion disks, and it is likely
that the two phenomena are related.  Jets provide accretion disks with an
efficient means to remove angular momentum, and disks provide jets with a
source of material and magnetic fields.  The magnetocentrifugal model has
been proposed \citep{bp82} as a way to generate a fast, collimated jet 
from an accretion disk.  In this model, gas flows away from
the accretion disk along open magnetic field lines that thread the disk.
The magnetic field accelerates the gas to super-fast magnetosonic speeds,
and the rotation of the gas winds the field up into a predominantly 
toroidal
configuration which is thought to be able to collimate the gas through 
``hoop'' stress (see, however, \citealt{ok99}).

A full analytical treatment of the steady flow is difficult because the
equation describing the cross-field force balance, the Grad-Shafranov
equation, changes its character from elliptic to hyperbolic upon crossing
the fast magnetosonic surface, the location of which is not known a priori.
In order to attack the problem, authors have had to make simplifying
assumptions.  \citet{bp82} presented an analytical solution for this flow
under the assumption of self-similarity in the original paper on
magnetocentrifugal jets, and self-similarity has also been employed by
others, e.g.\ \citet{cl94}, \citet{o97}, \citet{tts97} and \citet{fc04}.  
Another approach 
has been to make an assumption about the magnetic field configuration and 
then solve for the hydrodynamic quantities, e.g., 
\citet{pn86} and \citet{ks97}.  An
asymptotic analysis of the Grad-Shafranov equation can yield the structure
of the flow at large distances from the source, as in \citet{hn89, hn03}
and \citet{shu95}.

Because of the difficulties in finding analytical solutions to the flow 
structure, it has become common in recent years to attack the problem 
with time-dependent numerical simulations. 
Two different approaches have been taken in developing
numerical models of jet launching from accretion disks --- including the
accretion disk as part of the simulation, or treating the disk as a 
fixed boundary condition.  We have chosen the latter approach for our 
study and will focus on it.  We refer the reader to \citet{su86}, 
\citet{sn94}, \citet{m96}, and \citet{vonR04} for examples of models 
which include the disk.

For cases where the disk is treated as a fixed rotating boundary for the 
simulation,
it is important that the hydromagnetic quantities are specified on the disk
in a self-consistent manner.  Several different approaches have been taken
for specifying both the initial and boundary conditions for numerical
simulations of magnetocentrifugal jet launching.  \citet{u95} performed
simulations demonstrating the acceleration and collimation of a jet from
a disk with a hot (plasma $\beta$ parameter $\gg1$) corona. Their 
simulations did not reach steady-state, although
it is unclear whether this is a feature of their model or the result of
terminating the simulation before it became stationary.  A similar study
\citep{r97} using a stronger magnetic field and a cold ($\beta<1$) corona
resulted in stationary wind solutions; however the wind was at best poorly
collimated.  \citet{op97a,op97b} also use a corona that can be described
as cold, and two
different initial magnetic configurations:  a potential (current-free)
field \citep{op97a}, and a uniform, vertical (parallel to rotation axis)
field \citep{op97b}.  Both cases exhibit acceleration and collimation,
but the former comes to a steady-state, while the latter is characterized
by episodic behavior.  However, in a follow-up paper \citep{op99}, they
argue that it is not the initial magnetic configuration but rather the
mass loading at the base of the wind that determines whether a given model
will reach steady-state or display episodic behavior.  Their conclusions
are based on simulations they performed for different values of 
mass load, for both
the potential and uniform field configurations.  \citet{klb99}, hereafter
referred to as KLB99, also performed cold simulations in an initially
potential field, with an important modification.  They pinned the foot
points of the wind-launching field lines on the disk, but left the field 
lines free to vary in the radial and azimuthal directions both on the
disk and away from the disk in the wind zone. These degrees of freedom 
are needed to prevent sharp discontinuities from developing between 
the disk boundary and the magnetocentrigual wind 
(cf.\ \citealt{bt99} for the case of a spherical boundary). 

All the works mentioned in the preceding paragraph use a magnetic field
threading the entirety of the disk (i.e.\ the entire $z=0$ plane).  
However, a launching region that extends to the edge of the simulation 
box results in a portion of the flow reaching the boundary before having 
been accelerated to super-fast magnetosonic speeds.  As pointed out in 
KLB99, this has the effect of making the results of the simulation 
sensitive to the size and shape of the simulation box, which is 
undesirable. Therefore we have chosen to present simulations with a 
limited launching region in this paper, as was done in \citet{klb03}
(KLB03 hereafter). The size of the launching region
is left as a free parameter, which has the benefit of allowing us potentially
to model another proposed candidate for jet production, the
X-wind model \citep{shu94}. This model also uses the magnetocentrifugal
effect to accelerate and collimate a wind, but the required magnetic 
field, rather than coming from a wide area on the accretion disk (the 
so-called disk wind, see \citealt{kp00}), instead comes from a narrow 
region around $\varpi=R_{X}$, where $R_{X}$ is the point where the 
accretion disk is in co-rotation with the stellar magnetosphere.  

\subsection{Effects of Mass Loading} 

The focus of our investigation will be on the effects of mass loading 
on the structure of magnetocentrifugal winds. It is motivated in part 
by the observations of the outflows from young stellar objects (YSOs). 
\citet{bon96} established that deeply embedded Class 0 objects 
typically drive CO molecular outflows with momentum fluxes more than an 
order of magnitude higher than those of older Class I sources. If the 
molecular outflow is driven by a magnetocentrifugal wind from a region 
close to the central object in a momentum conserving fashion, as 
currently believed, then the wind should have a higher mass flux 
in the Class 0 phase, since its wind speed can hardly be more than an 
order of magnitude faster; if anything, the fact that Class 0 sources 
are yet to accrete most of their final masses points to a smaller 
Keplerian speed in the wind-launching region, which would argue 
for a lower wind speed. The decrease in mass flux as the wind ages 
is probably correlated with the decrease in disk accretion rate. The 
same trend appears to continue into the optically revealed, classical 
T Tauri phase, where the primary wind can be probed directly in a 
number of ways \citep{erm93}. From the early embedded to
late revealed phase, the mass loss rate in a typical YSO wind may span 
a large range, from $\sim 10^{-6}\solarmassyear$ \citep{bon96} 
to $\sim 10^{-9}\solarmassyear$ or less \citep{ed93}. The 
question is then: how does the mass loading affect the structure of 
the wind? 

Part of the answer can be obtained analytically, by considering the 
wind properties along a flux tube of prescribed opening. \citet{s96} 
examined the special case of a cold wind in a radial field geometry. 
He introduced a dimensionless parameter which, for a non-radial
geometry, is equivalent to 
\begin{equation}
\mu\equiv {4\pi \rho v_p \varpi\Omega \over B_p^{2}}, 
\label{para1}
\end{equation}
where the mass density $\rho$, poloidal flow velocity $v_p$, cylindrical 
radius $\varpi$, angular speed $\Omega$, and poloidal field strength $B_p$ 
are evaluated at the base of the wind. The parameter measures the ability 
of the rotating magnetic field $B_p$ to accelerate the mass load, which 
is proportional to $\rho v_p$. We will refer to $\mu$ as the mass loading
parameter. In the ``light'' wind limit $\mu \ll 1$, 
\citet{s96} found that the flow is accelerated centrifugally along
the more or less rigid field line, up to the Alfv\'en radius $\varpi_A$,
which is located about $(3/2)^{1/2}\mu^{-1/3}$ times the foot point 
radius. In the opposite, ``heavy'' wind limit $\mu\gg 1$, the Alfv\'en 
radius moves to $(3/2)^{1/2}$ times the foot point radius. The field 
lines become tightly wound, and the flow is accelerated mainly by 
magnetic pressure gradients. In both parameter regimes, the terminal 
wind speed is given by 
\begin{equation}
v_{\infty}=\mu^{-1/3} v_0,
\label{terminal}
\end{equation}
where $v_0=\Omega\varpi$ is the rotational speed at the launching surface. 
The above relation results from the fast magnetosonic point being at 
infinity for a radially diverging flux tube, as in
\citeauthor{m69}'s (\citeyear{m69})
minimum energy solution for cold relativistic MHD winds \citep{gj70}.
\citet{ks97} showed that the above relation 
holds approximately, at least for light winds, even when the prescribed
field geometry is non-radial, as long as the fast point (which is now
at a finite radius) is not too close to the foot point. We therefore 
expect no major surprises for the wind acceleration, which is mainly 
determined ``locally'' along a field line by the conserved quantities
(see equations [\ref{eqn:kappa}]--[\ref{eqn:e}] below).

The effects of mass loading on the wind collimation are more difficult 
to determine. Collimation is a global wind property controlled by the 
cross-field force balance. One must solve the Grad-Shafranov equation, 
which has not been possible in general until recently. In this paper, 
we will 
use the MHD code developed in KLB03 to solve for the steady-state wind 
structure through time-dependent simulations. The goal is to study the 
effects of mass loading on both wind collimation and acceleration. We
find that, as expected, a heavier loading leads to a better collimation
and a slower wind, and that the slowdown can be described to a good 
approximation by equation (\ref{terminal}). We also show that there 
exists a maximum mass load for a given wind-launching magnetic field, 
beyond which the magnetocentrifugal mechanism shuts off, and that the 
maximum corresponds to the dimensionless parameter $\mu\gsim 1$, with 
the exact value depending on the injection velocity at the launching
surface. We show that, despite significant collimation in the wind,
some of its most important properties can still be well-described by the
analytic results originally derived for a wind in a radial field geometry. 

The remainder of the paper is organized as follows.  In \S\,\ref{desc}
we describe 
our formulation of the magnetocentrifugal wind problem and the setup 
of numerical simulation. Our reference 
simulation is presented in detail in \S\,\ref{reference}.  We compare
models with different mass loads and distributions of mass loading in
\S\,\ref{loadmagn} and \S\,\ref{loaddist} respectively. 
A discussion of these results, along with our conclussions, are given in
\S\,\ref{conclusion}.

\section{FORMULATION OF THE PROBLEM}
\label{desc}

\subsection{Governing Equations}

We consider a system consisting of a central gravitating mass surrounded by
an accretion disk threaded with a magnetic field. The system's axisymmetry
suggests a cylindrical coordinate system ($z$, $\varpi$, $\phi$) with the
central mass situated at the origin, the accretion disk lying in the $z=0$
plane, and the axis of rotation along the $\varpi=0$ axis.

Our interest is in finding steady-state wind solutions for this model 
through time-dependent numerical simulations. The standard MHD wind
equations are
\begin{eqnarray}
{\partial \rho\over \partial t}+\nabla\cdot(\rho\mathbf{v})&=&0,\\
\rho {\partial\mathbf{v}\over \partial t}+\rho(\mathbf{v}\cdot\nabla)
\mathbf{v}
&=&-\nabla p-\rho\nabla\Phi_{g}+\frac{1}{4\pi}\,
(\nabla\times\mathbf{B})\times\mathbf{B} , \\
{\partial \mathbf{B}\over \partial t}&=& \nabla\times(\mathbf{v}
\times\mathbf{B}) , \\
\nabla\cdot\mathbf{B}&=&0 ,
\end{eqnarray}
where $\mathbf{B}$ is the magnetic field, $\mathbf{v}$ is the velocity field,
$\rho$ is the mass density, $p$ is the thermal pressure, and $\Phi_{g}$ is 
the gravitational potential.  

It is well known that, in steady state, there are four conserved quantities 
along each magnetic field line \citep{m68}:
\begin{eqnarray}
\label{eqn:kappa}
\kappa&=&\frac{\rho v_{p}}{B_{p}},\\
\label{eqn:omega}
\Omega&=&\frac{1}{\varpi}\Big(v_{\phi}-\frac{B_{\phi}}{B_{p}}
v_{p}\Big),\\
\label{eqn:l}
L&=&\varpi\,\Big( v_{\phi}-\frac{B_{\phi}}{4 \pi \kappa}\Big),\\
\label{eqn:e}
E&=&\frac{v^{2}}{2}+h+\Phi_{g}-\frac{B_{\phi}B_{p}\Omega\varpi}
{4\pi\rho v_{p}}.
\end{eqnarray}
These can be interpreted respectively as the conservation along magnetic
field lines of mass to magnetic flux ratio, angular velocity, specific
angular momentum, and specific energy.  Here $h$ is the specific enthalpy
and the subscript $p$ indicates a quantity in the poloidal ($z$, $\varpi$) 
plane. In our simulations, $\kappa$ and $\Omega$ are prescribed, while 
$L$ and $E$ are to be determined numerically.   

\subsection{Boundary and Initial Conditions}

All of our simulations have been performed using the Z{\scriptsize EUS}3D 
MHD code \citep{cnf94}, with modifications as described 
in KLB99 and KLB03.  Most of 
the modifications are related to boundary and initial conditions, which 
we describe briefly below. 

The outer boundaries of the simulation box at $z=z_{\rm{max}}$
and $\varpi= \varpi_{\rm{max}}$ use the standard outflow
boundary conditions present in the Z{\scriptsize EUS}3D code.
Specifically the values of all the variables in the ghost zones 
are set equal to the the values at $z=z_{\rm{max}}$ or
$\varpi=\varpi_{\rm{max}}$. The axial boundary $\varpi=0$ is
handled with standard reflection boundary conditions, i.e.\ the ghost
zone values of 
 the variables are reflections of the simulation zone quantities with a 
sign change in the $\varpi$ and $\phi$ (but not $z$) 
components of vector quantities.
The $z=0$ boundary is the most problematic of the boundaries to implement.
We have divided the $z=0$ surface into two regions:  an inner launching
surface and an outer surface along which the plasma loaded onto the
last field line slides. For the region interior to the 
maximum launching radius $\varpi_{0}$, we pin the field lines at their 
foot points, but allow them to bend freely in the radial and azimuthal 
directions. This is accomplished through imposing conditions on the 
electromotive force field $\mathbf{\cal E}$ in the ghost zones (KLB99). 
Exterior to $\varpi_{0}$, we demand that the last field line to lie
exactly on the equator (so that $v_z=B_z=0$). The requirement is 
enforced through ${\cal E}
_{\phi}(-z)=-{\cal E}_{\phi}(z)$, ${\cal E}_{\varpi}(-z)=-{\cal E}
_{\varpi}(z)$, and ${\cal E}_{z}(-z)={\cal E}_{z}(z)$. For the initial
distribution of magnetic field in the active zones of the simulation 
box, we adopt a 
potential configuration computed using the prescribed magnetic flux 
distribution on the launching surface as a boundary condition. The 
computation method is given in KLB03. The potential field is extended
into the ghost zones by solving the Laplace equation for the magnetic
flux function.  We fill the computational domain
with a low-density ambient medium, which is completely replaced by
the wind material coming off the launching surface well before the end 
of the simulation. 

\subsection{Simulation Setup} 

As a lower boundary to our wind simulation, we consider an infinitely thin 
disk around a central star, idealized as 
a point mass $M_*$ at the origin. To avoid singularity, we soften the 
stellar gravitational potential $\Phi_g$ within a spherical radius $r=
\varpi_g$, according to 
\begin{equation}
\Phi_{g}(r)=\left\{\begin{array}{ll}-\frac{G M_{*}}{\varpi_{g}}
\Big(1.2 - 0.2(\frac{r}{\varpi_{g}})^{5}\Big) & , r\leq\varpi_{g}\\
-\frac{G M_{*}}{r} & ,r >\varpi_{g}\mbox{.}\end{array}\right.
\end{equation}
To be in a mechanical equilibrium, the disk must rotate at a speed
\begin{equation}
v_d(\varpi)=\sqrt{\varpi {d\Phi_{g}(\varpi)\over d\varpi} }
=\left\{\begin{array}{ll}\left({G M_*\over \varpi_g}\right)^{1/2}
\left({\varpi\over \varpi_g}\right)^{5/2} & ,
\varpi\leq\varpi_{g}\\
\left({G M_*\over \varpi}\right)^{1/2} & ,
\varpi_{g}<\varpi<\varpi_{0}\mbox{.}\end{array}\right.
\end{equation}

>From the disk surface, we inject cold material of negligible thermal 
pressure into the wind supersonically, with an initial vertical speed 
given by 
\begin{equation}
v_{z}(\varpi)=\left\{\begin{array}{ll}V_{i} \sqrt{-\Phi_g(\varpi)} & ,
\varpi\leq\varpi_{g}\\
V_{o}v_d(\varpi)\,{\cal S}(\varpi) & ,
\varpi_{g}<\varpi<\varpi_{0}\mbox{.}\end{array}\right.
\label{initvel}
\end{equation}
The dimensionless 
parameter $V_{i}$ ($\ge 1$) controls the injection speed inside the 
softening radius where the magnetocentrifugal mechanism is ineffective,
\footnote{Even though a steady magnetocentrifugal wind is impossible 
in this region, it may be filled with outflows driven by other 
mechanisms, such as a stellar wind or blobs driven 
impulsively by magnetic reconnection events (e.g., Hayashi, Shibata 
\& Matsumoto 1996).} 
and $V_{o}$ ($\ll 1$) is for the wind material to be accelerated 
magnetocentrifugally from the Keplerian disk between $\varpi_g$ and 
$\varpi_0$. The spline function 
\begin{equation}
{\cal S}(\varpi)=\sqrt{1-\Big( \frac{\varpi-\varpi_{g}}
{\varpi_{0}-\varpi_{g}}\Big)^{2}}
\end{equation}
is chosen to bring $v_z$ continuously to zero at the edge of the launching 
region, as required by our boundary conditions. 

We adopt a softened power-law form for the distribution of the 
vertical component of magnetic field $B_{z}(\varpi)$ on the launching surface
\begin{equation}
B_{z}(\varpi)=\left\{\begin{array}{ll}B_{0}\Big(1.2-0.2(\frac{\varpi}
{\varpi_{g}})^{5}\Big)^{\alpha_{B}} & ,\varpi\leq\varpi_{g}\\
B_{0}\Big(\frac{\varpi}{\varpi_{g}}\Big)^{-\alpha_{B}}{\cal S}(\varpi) 
& , \varpi_{g}<\varpi<\varpi_{0}\mbox{,}\end{array}\right.
\label{fieldstrength}
\end{equation}
where the parameter $B_{0}$ controls the strength of the magnetic field,
and $\alpha_{B}$ the spatial distribution. It is also smoothed by the 
spline function ${\cal S}(\varpi)$. To complete the specification of 
the launching conditions, we adopt the following form for the mass density 
distribution
\begin{equation}
\rho(\varpi)=\left\{\begin{array}{ll} {D_0 \over 2 v_z(\varpi)} 
& ,\varpi\leq\varpi_{g}\\
{D_0\over V_o v_d(\varpi)}\left({\varpi\over \varpi_g}\right)
^{-\alpha_m} & , \varpi_{g}<\varpi<\varpi_{0}\mbox{,}\end{array}\right.
\label{density}
\end{equation}
where the parameter $D_0$ controls the rate of mass loading and $\alpha
_m$ the spatial distribution. 

All of our calculations within Z{\scriptsize EUS}3D are carried out in
dimensionless units for convenience, but it is instructive to
redimensionalize the hydromagnetic quantities at the end of the simulation
for comparison with observational results.  For our application to YSOs,
we set the stellar mass $M_{*}= 1\solarmass$ and the softening radius
$\varpi_{g}=0.1\AU$, which yield a characteristic velocity scale, the 
Keplerian velocity at $\varpi_{g}$, $\sqrt{GM_{*}/\varpi_{g}}=94\,\kms$. 
The scales for other quantities can be determined once the magnetic 
field strength at the gravitational softening radius, $B_0$, is specified.

\section{RESULTS}
\label{results}

We begin with a detailed description of our reference run in \S\, 
{\ref{reference}}, to which the rest of our simulations are compared.  
In \S\,\ref{loadmagn} we present a
series of simulations in which only the total mass loading of the 
wind, $\dot{M}_{w}$, is varied from the reference run.  Finally in 
\S\,\ref{loaddist} we present a smaller group of simulations where 
the total mass load is fixed, but its distribution over the 
launching surface varies.

\subsection{Reference Wind Solution} 
\label{reference}

For our reference simulation, we have chosen a size for the 
launching region
of ten times the gravitational softening radius: $\varpi_0=10 \varpi_g
=1\AU$. Inside $\varpi_g$, the dimensionless injection parameter is set 
to $V_i=2$, so that the injected material can escape without the help 
from the magnetocentrifugal effect. For the launching region that rotates  
in a Keplerian fashion (between $\varpi_g$ and $\varpi_0$), we let $V_o
=0.01$, so that the injection speed is much less than the local Keplerian 
speed (but still greater than the sound speed, which is set to an arbitrarily
small value). For the density distribution at the base of the wind, we set 
$\alpha_m=2$, which is the same as that adopted by \citet{bp82}
for their self-similar solutions, and we choose $D_0=0.1$. 
To specify the wind-launching magnetic field, we adopt a vertical 
field strength at $\varpi_g$ of $B_0=19.2\gauss$ and an exponent for
field distribution, $\alpha_B=5/4$. The latter is again the Blandford-Payne 
scaling. The adopted field strength leads to a wind mass loss rate of 
$\dot{M}_{w}=10^{-8}\solarmassyear$ from each side of the disk. 
The corresponding scale for mass density at $\varpi_g$ is 
$2.07\times 10^{-14}\massden$, or, assuming pure 
hydrogen gas, a number density scale of $1.23\times10^{10}\numden$. For other
choices of $B_0$, the mass flux and density scale vary as 
$B_0^2$. 

\begin{figure*}
        \begin{center}
                \epsscale{0.55}
                \plotone{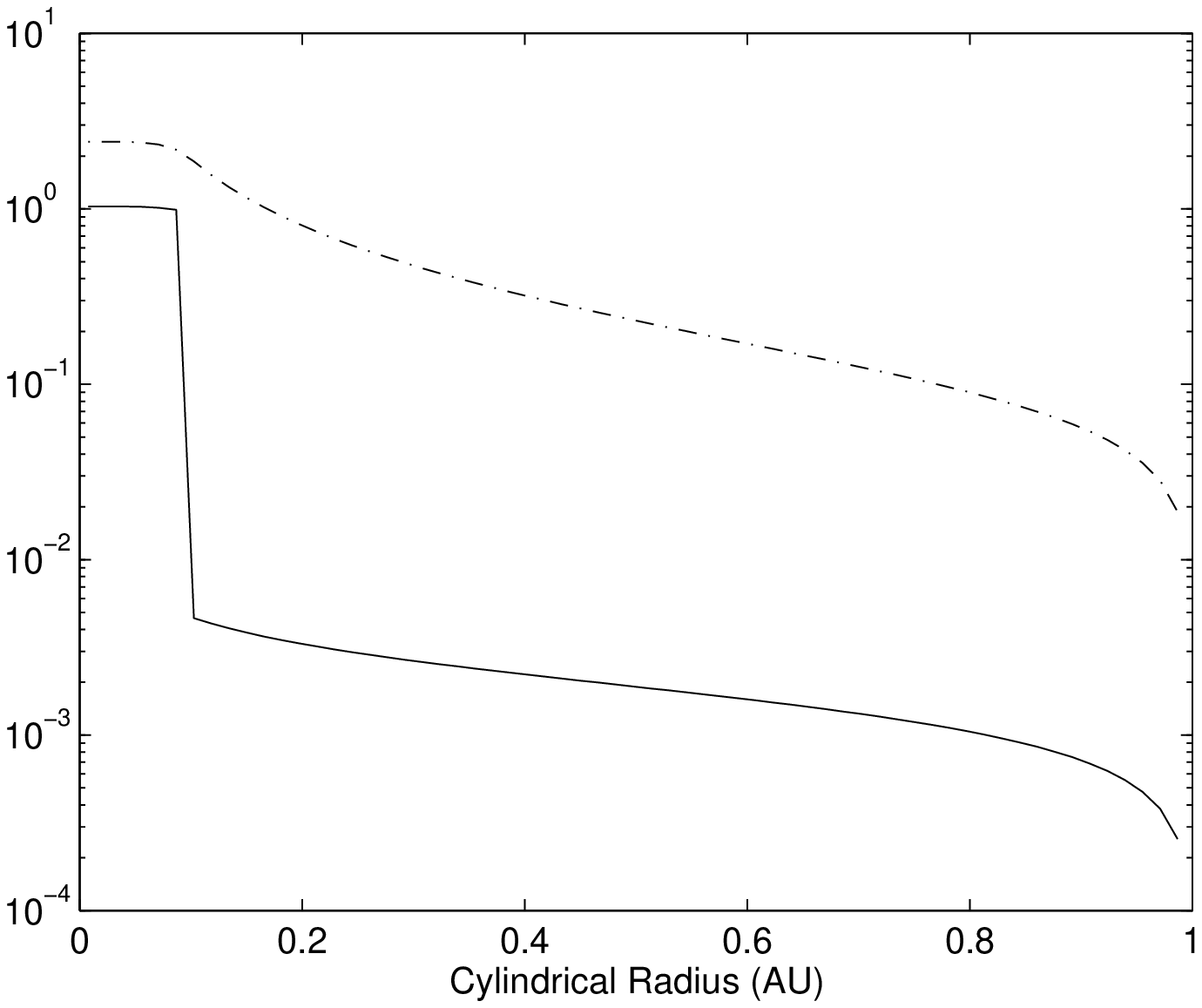}
                \plotone{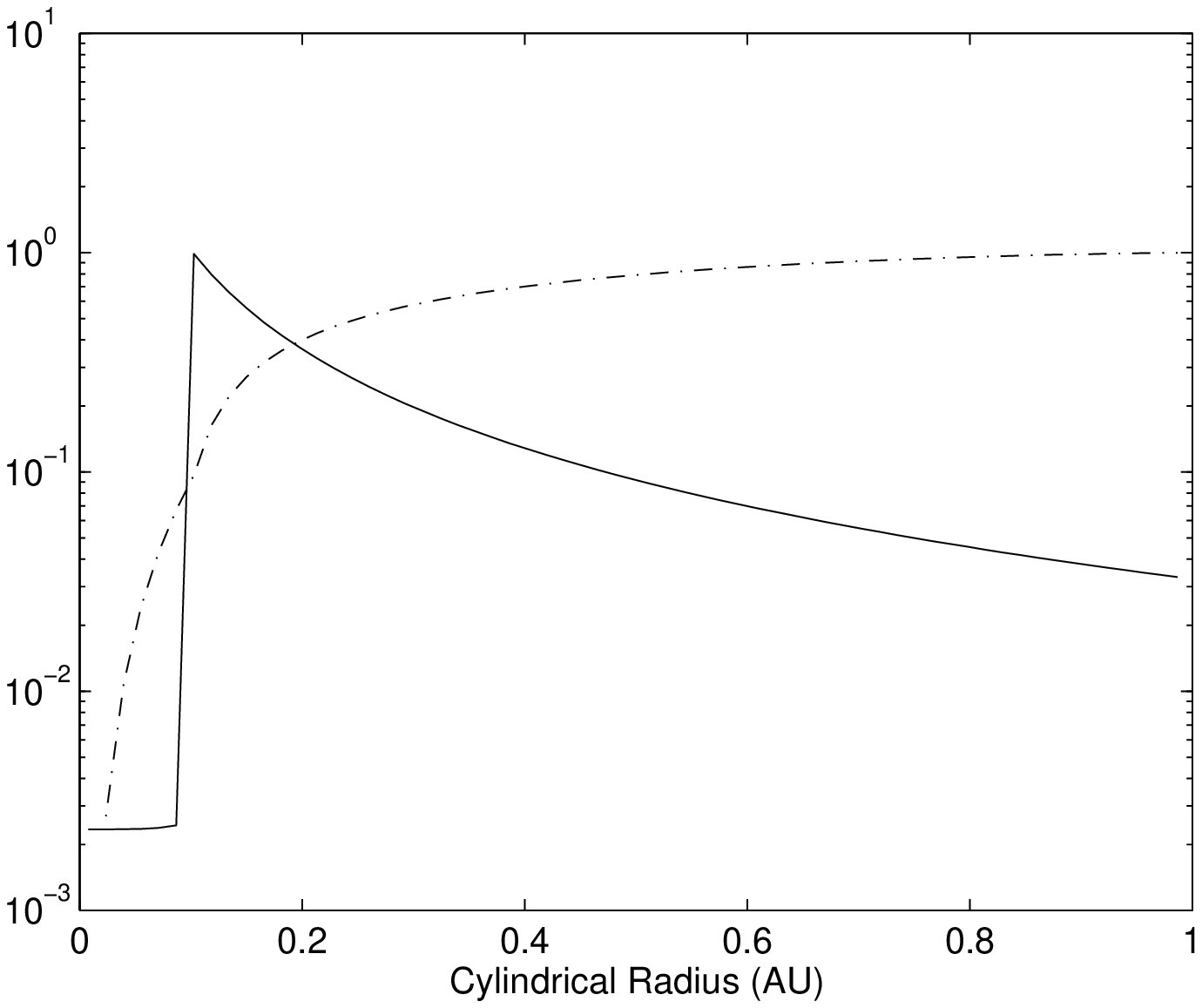}
        \end{center}
        \caption{Launching conditions of the reference simulation. Plotted are
the vertical component of the injection velocity in units of $200\kms$ 
(solid) and the vertical field strength in units of $20\gauss$ (dash-dot)
in the upper panel, and   
the mass density at the base of the wind in units of $2 \times 10^{-13}\massden$
(solid) and the cumulative mass flux from each side of the disk in units of
$10^{-8}\solarmassyear$ (dash-dot) in the lower panel.}
        \label{fig:f1}
\end{figure*}

Figure \ref{fig:f1} shows the prescribed launching conditions for our 
reference simulation. 
Note that the axial injection region within 
$\varpi_g=0.1\AU$ contains about 10\% of the cumulative mass flux 
from the disk in our simulation. This fraction can be made smaller
by reducing the injection density, but it would take longer for 
the wind to reach a steady state because of a more stringent Courant 
condition. KLB03 investigated the effects of the axial injection 
and concluded that the structure and dynamics of the magnetocentrifugal 
part of the wind remained largely unchanged for differing mass fluxes 
in the axial region. We confirmed this result with a new set of
simulations. 

Computationally, we have used a grid with 256 active zones in both the 
$\varpi$ and $z$ directions. On both axes the grid spacing is linear 
for $0\leq\varpi,z\leq1.2\AU$ with 76 zones, and logarithmic for $1.2
\leq\varpi,z\leq100\AU$ with 180 zones. This arrangement allows us to 
extend our simulation to large distances from the origin while still 
retaining good resolution of the launching region.

\begin{figure*}
        \begin{center}
                \epsscale{0.55}
                \plotone{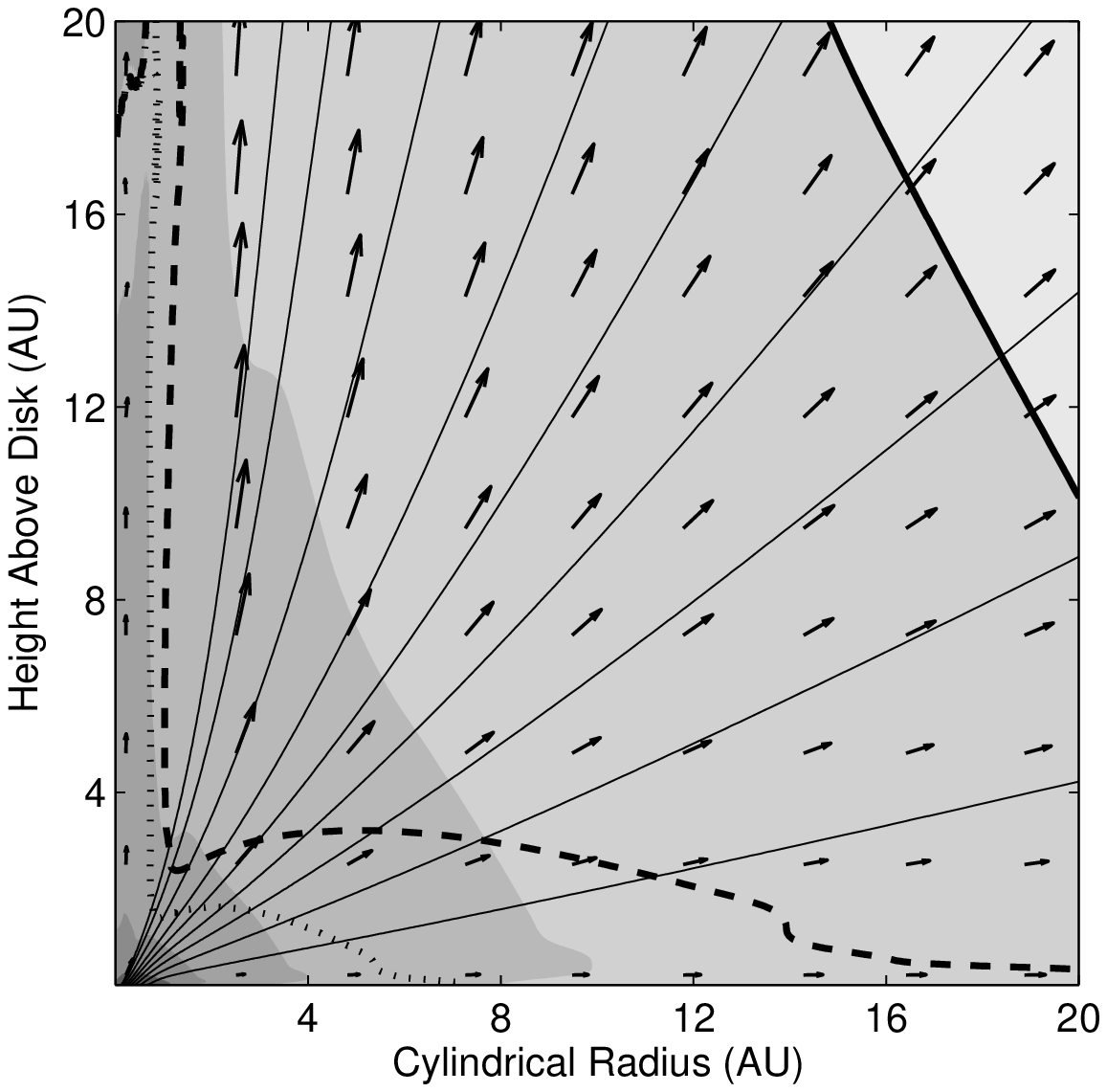}
                \plotone{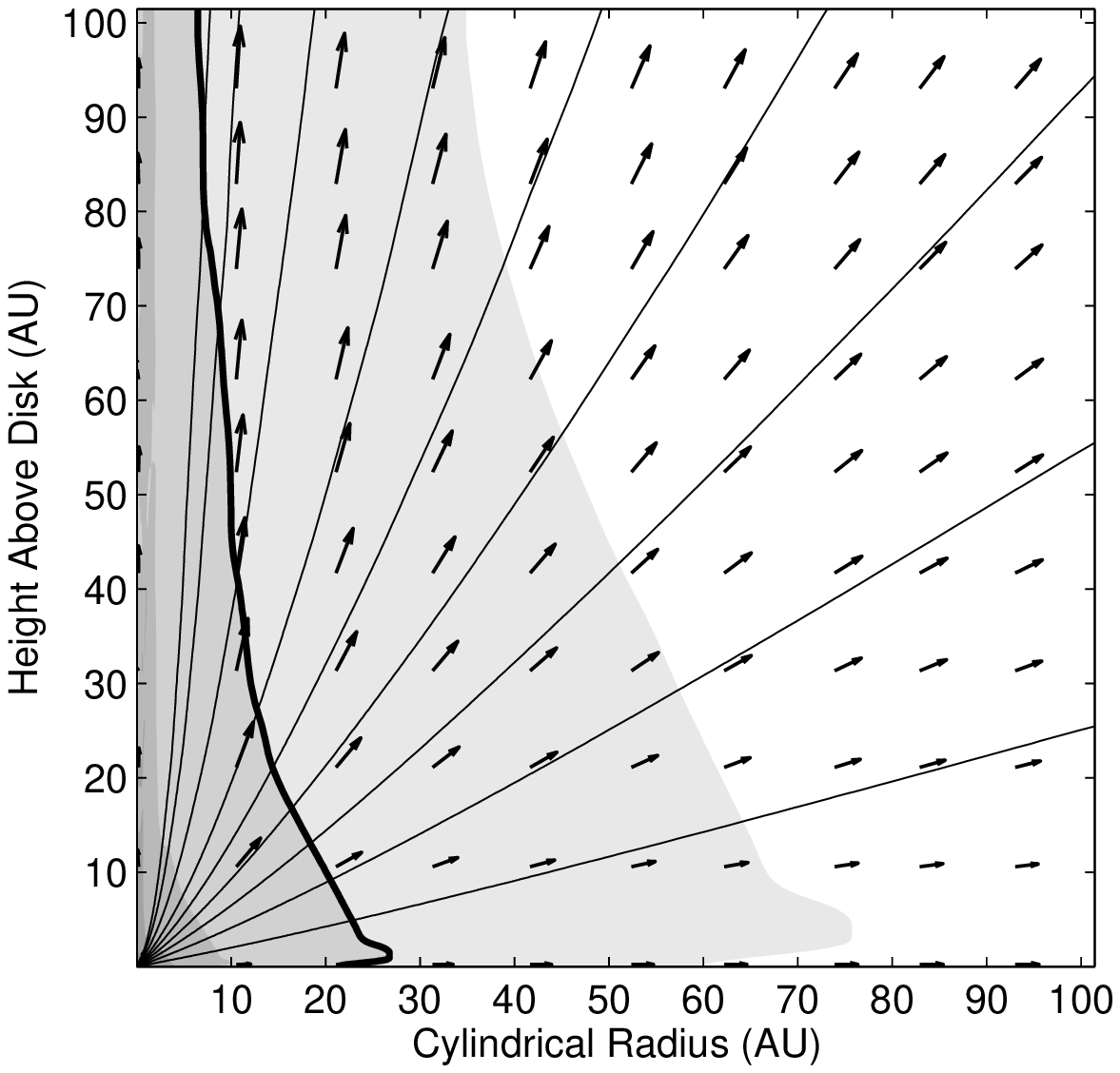}
        \end{center}
        \caption{Steady-state wind solution in the reference simulation. 
Plotted as thin solid lines are the nine magnetic field lines that divide
the wind into ten zones of equal mass flux.  The velocity vectors are denoted
by arrows, with the length of the arrow proportional to the poloidal flow speed.
Density contours are shown in shades (one decade per shade), with the thick
solid line marking $n_H=10^4\numden$.  The Alfv\'{e}n and fast magnetosonic
surfaces are shown in the top panel as a thick dotted line and thick dashed
line respectively.  The top panel shows the inner region of the simulation,
while the lower one shows the full simulation box.} 
        \label{fig:f2}
\end{figure*}

\begin{figure*}
        \begin{center}
               \epsscale{1.0}
               \plotone{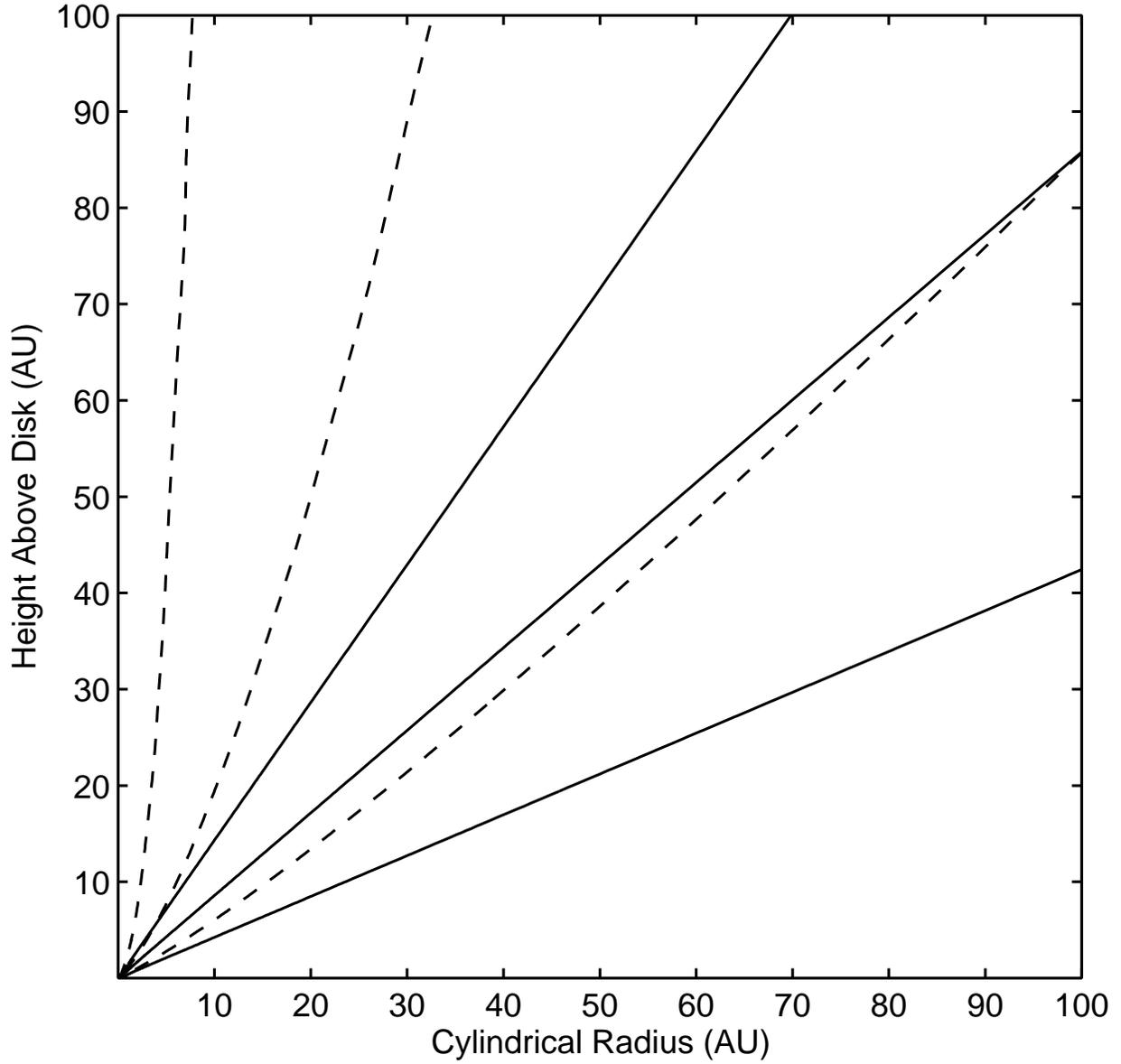}
        \end{center}
        \caption{Comparison of initial to final magnetic field 
configuration for field lines enclosing 25\%, 50\%, and 75\% of the
 total mass flux from the launching surface.  The initial potential
 field is shown in solid lines, while the final, steady-state, 
field is shown in dashed lines.}
        \label{fig:f3}
\end{figure*}

The results of our reference simulation are shown in Figures
\ref{fig:f2}--\ref{fig:f4}. In 
Fig.\,\ref{fig:f2}, we plot the steady-state solution on two scales. On 
the smaller, $10\AU$ scale, we find that most of the space is filled 
with magnetocentrifugally accelerated wind material from the Keplerian
part of the launching surface between $0.1$ and $1\AU$. The fast injection 
part of the wind, 
enclosed within the streamline closest to the axis, occupies a relatively
small fraction of the space. The fraction decreases on the larger scale, 
as shown in the bottom panel. There is some residual nonsteadiness 
in the fast injection region, which has little effect on the magnetocentrifugal
part of the wind. In both panels, gradual collimation is 
evident in both field line and density contour. The collimation appears
more prominent in the axial region than in the equatorial region. Note 
the presence of bulging in density contours at low altitudes above the 
disk. The degree of bulging is related to the slope of the mass loading 
$\alpha_{m}$, as we show in \S\,\ref{loaddist} (see also KLB03). It may
affect the appearance of the axial jet. 

Before discussing more quantitatively the acceleration and collimation 
of the reference wind solution, we note that its dimensionless mass 
loading parameter at the Keplerian part of the launching surface is 
given by
\begin{equation} 
\mu(\varpi_g\le \varpi \le \varpi_0)=\mu_g \left({\varpi\over\varpi_g}
\right)^{2\alpha_B-\alpha_m-1/2} \left[{B_z(\varpi)\over B_p(\varpi)
{\cal S}(\varpi)}\right], 
\label{mu}
\end{equation}
where the scaling constant 
\begin{equation}
\mu_g\equiv {4\pi D_0\over B_0^2}\left({GM_*\over \varpi_g}\right)^{1/2}
=6.25\times 10^{-3}
\label{mu_g1}
\end{equation}
(where the density and field strength constants $B_0$ and $D_0$ are 
defined in equations  [\ref{fieldstrength}] and [\ref{density}]) and 
the exponent $2\alpha_B-\alpha_m-1/2=0$ for the reference run. 
The combination in 
the square bracket is not predetermined, since $B_z/B_p=\cos(\theta)$
and the angle $\theta$ between the poloidal field line and the disk
normal is allowed to vary in response to the stresses in the wind. We
find that in steady state the combination has values of order unity 
or less. Therefore, the
reference wind has $\mu(\varpi)\ll 1$ on the launching surface. It is
a ``light'' wind.  

Even though the mass loading is relatively light, it causes significant 
flow collimation. This is shown in Fig.\,\ref{fig:f3}, where three
representative 
field lines enclosing respectively 25, 50 and $75\%$ of the total mass 
flux are drawn. Also shown for comparison are the initial positions
of these field lines, which correspond to the potential (current-free) 
field outside the launching surface; they mark where the field lines 
would be in the absence of the wind. Evidently, there is enough cross-field
electric 
current flowing in the wind to substantially modify the initial potential 
field configuration, particularly at large distances, where the magnetic
field is dominated by the toroidal component. 

The flow collimation enables flux tubes to diverge faster than radial,
causing the fast magnetosonic point to move from infinity to a finite
radius (see the location of the fast surface in Fig.\,\ref{fig:f2}). Beyond
the fast
surface, flow acceleration continues, until most of the magnetic energy 
is converted into the kinetic form. The efficient conversion can be 
seen in the upper panel of Fig.\,\ref{fig:f4}, where the fast 
magnetosonic Mach 
number $M_f=\sqrt{4\pi\rho v_p^2/(B_p^2+B_\phi^2)}$ is plotted as a 
function of spherical radius for the three 
selected field lines. The wiggles on the curve for the $25\%$ mass flux
enclosing field line are caused by the nonsteady flow near the axis. 
At large distances away from the launching region, 
the ratio of kinetic to magnetic energy is approximately $M_f^2/2$ 
\citep{s96}. This ratio ranges from 2 to 3.1 at a radius of $100\AU$.
They are much larger than the 
ratio of $\sim 0.5$ at $M_f=1$, pointing to significant acceleration 
of the wind plasma beyond the fast surface. 

The lower panel of Fig.\,\ref{fig:f4} displays the pitch angle
$\theta_{B}=\tan^{-1}(\vert B_{\phi}\vert /B_{p})$
of the magnetic field along the three representative 
field lines. It shows clearly that the wind-launching field lines start 
out more or less straight (i.e., having small pitch angle), in accordance 
with the wind being relatively light. They become increasingly toroidal 
at large distances, especially beyond the Alfve\'n surface. The toroidal 
magnetic pressure gradient dominates the wind acceleration beyond the 
fast point (see also Fig.\,\ref{fig:f13} below). Inside the fast point the 
centrifugal effect dominates. We 
now show that this basic behavior persists as long as the mass loading 
is lower than some critical amount. 

\subsection{Variation in Total Mass Loading}
\label{loadmagn}

To explore the effects of mass loading on the wind structure, we have 
performed a series of simulations in which the only parameter that 
was varied from the reference run was the total amount of mass loading
${\dot M}_w$. We were able to
obtain reliable solutions over a wide range of ${\dot M}_w$, from 
$0.1$ to $3000$ times the value for the reference solution,
corresponding to $1\times 10^{-9}$ -- $3\times 10^{-5}$ in units of
$(B_0/19.2\gauss)^2\solarmassyear$. The mass loading used for each
simulation (assuming the fiducial choice $B_0=19.2\gauss$), as well as 
some important data from the results, are summarized in Table 
\ref{tab:t1}.  The second 
column shows the value of the characteristic mass loading parameter 
$\mu_g$ on the launching surface that is defined in equation (\ref{mu_g1}).  
The quantities in the last three columns, $v_{\infty}$, $M_{f,\infty}$, 
and $\varpi_j$, are measured at a spherical radius of $100\AU$ along 
the 50\% mass flux enclosing field line, which comes from the same
location on the launching surface for all cases. 
As expected, the wind becomes 
slower and more collimated as the mass loading increases.  Half of
the mass flux is enclosed within $14\AU$ of the axis for the $3\times 
10^{-5}\solarmassyear$ wind, whereas the same fraction is enclosed 
within $48\AU$ in the $1\times 10^{-9}\solarmassyear$ case. 

\begin{figure*}
        \begin{center}
               \epsscale{1.1}
               \plotone{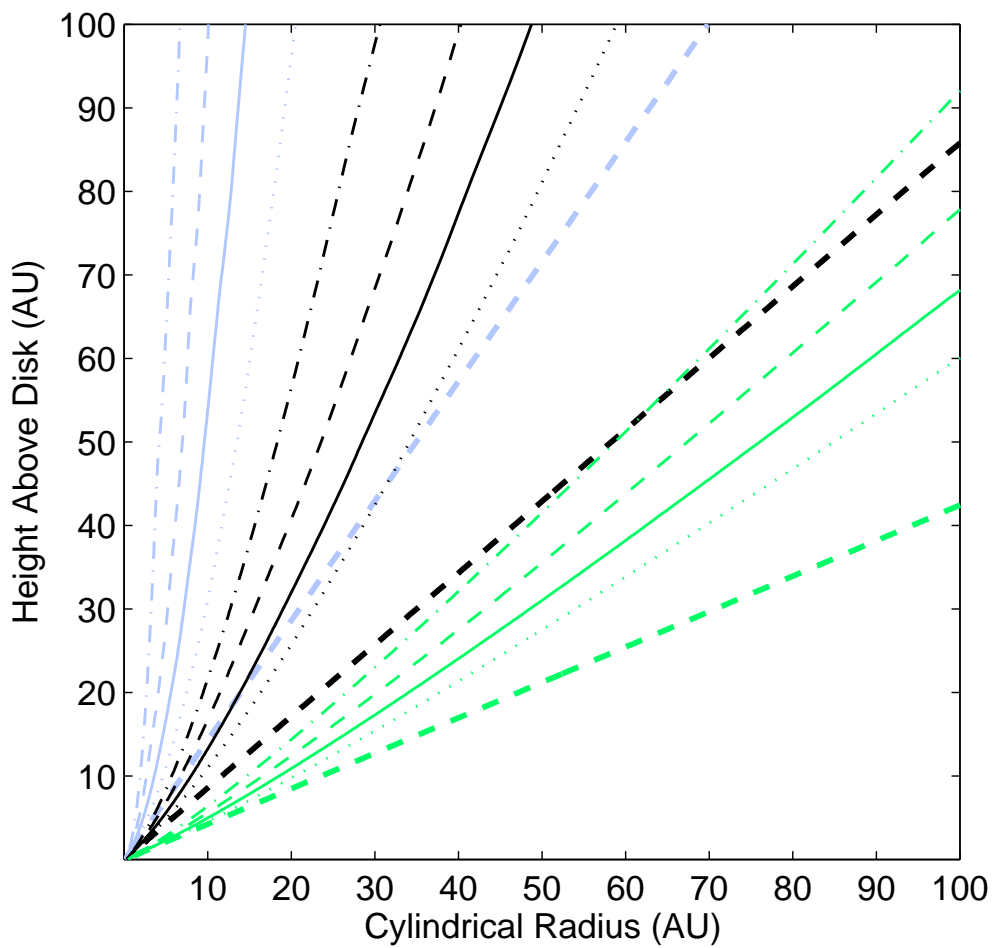}
        \end{center}
        \caption{The 25\% (blue), 50\% (black), and 75\% (green) mass flux 
enclosed field lines plotted for the simulations with $\dot{m}_0 = 0.1$
(dotted lines), $\dot{m}_0=1$ (reference solution, solid), 
$\dot{m}_0=10$ (dashed), and $\dot{m}_0=100$ (dash-dotted).  The degree 
of magnetic field collimation clearly 
increases with mass load. Also plotted in thick dashed lines
is the initial potential field.}
        \label{fig:f5}
\end{figure*}

To illustrate the wind collimation in more detail, we plot in 
Fig.\,\ref{fig:f5} 
three field lines enclosing respectively $25$, $50$ and $75\%$ of 
the total mass flux for each of the four representative cases ${\dot 
m}_0=0.1$, $1$, $10$, and $100$, where ${\dot m}_0$ is the wind mass flux 
${\dot M}_w$ scaled by $10^{-8}\times (B_0/19.2\gauss)^2\solarmassyear$; 
the ${\dot m}_0=1$ case thus corresponds to the reference run. The 
scaled mass flux is related to the characteristic mass loading 
parameter defined earlier by 
\begin{equation}
\mu_g=6.25\times 10^{-3}{\dot m}_0.
\label{mu_g2} 
\end{equation}
Note that the lightest (${\dot m}_0=0.1$) wind is also the least 
collimated, as expected. 
Nevertheless, its field lines are still much better collimated than
the initial potential configuration, despite the small dimensionless 
mass loading parameter at the launching surface: $\mu \sim 5\times 
10^{-4}$. As ${\dot m}_0$ is lowered even further, 
the fast surface starts to extend beyond the computation box, making 
the simulation unreliable. Furthermore, the Courant condition becomes
more stringent for the lower values of mass loading, setting a 
practical lower bound on ${\dot m}_0$ because of computer time constraints.
In any case, judging from the spacing between 
the field lines of different mass loading in Fig.\,\ref{fig:f5}, it
appears that 
a further reduction of ${\dot m}_0$ by at least several orders of magnitude 
would be needed for the field lines to approach the potential 
configuration. An implication is that the potential  
configuration can rarely, if ever, be a good approximation to the 
magnetic field struture of the (non-relativistic) magnetocentrifugal 
winds relevant for star formation, particularly at large distances; 
the structure must be determined numerically by solving the cross-field 
force balance equation.  

\subsubsection{Steady ``Light'' Winds}

\begin{figure*}
        \begin{center}
                \epsscale{1.0}
                \plottwo{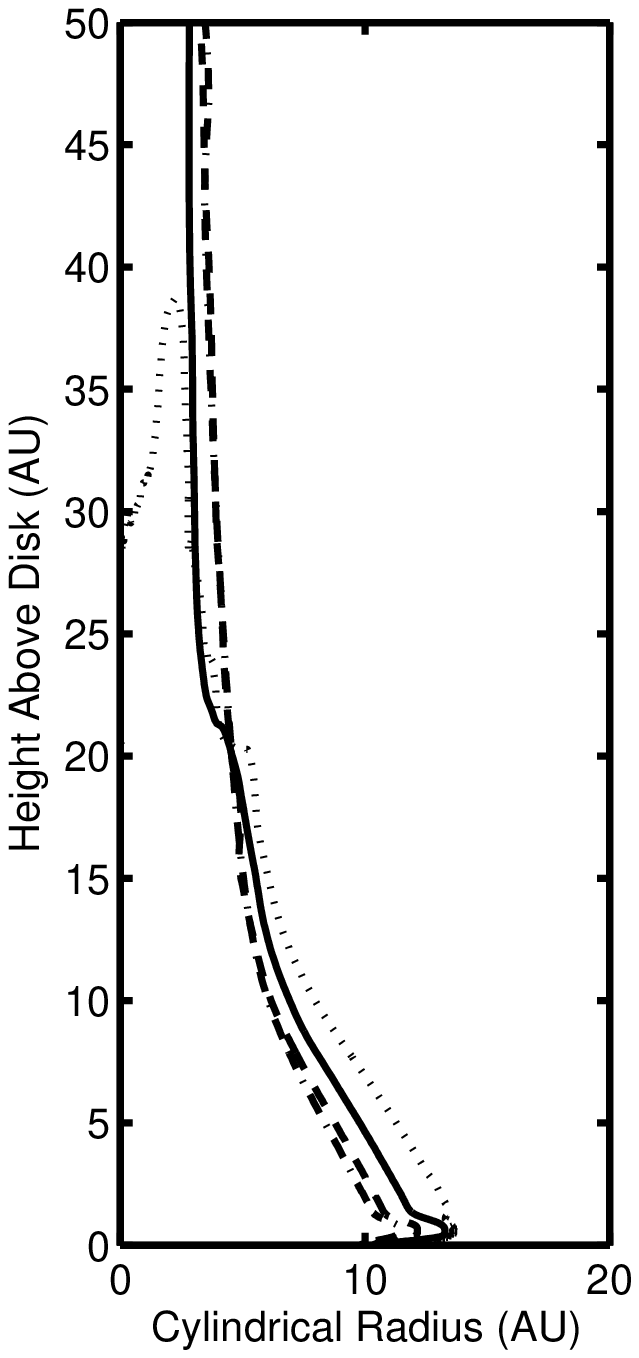}{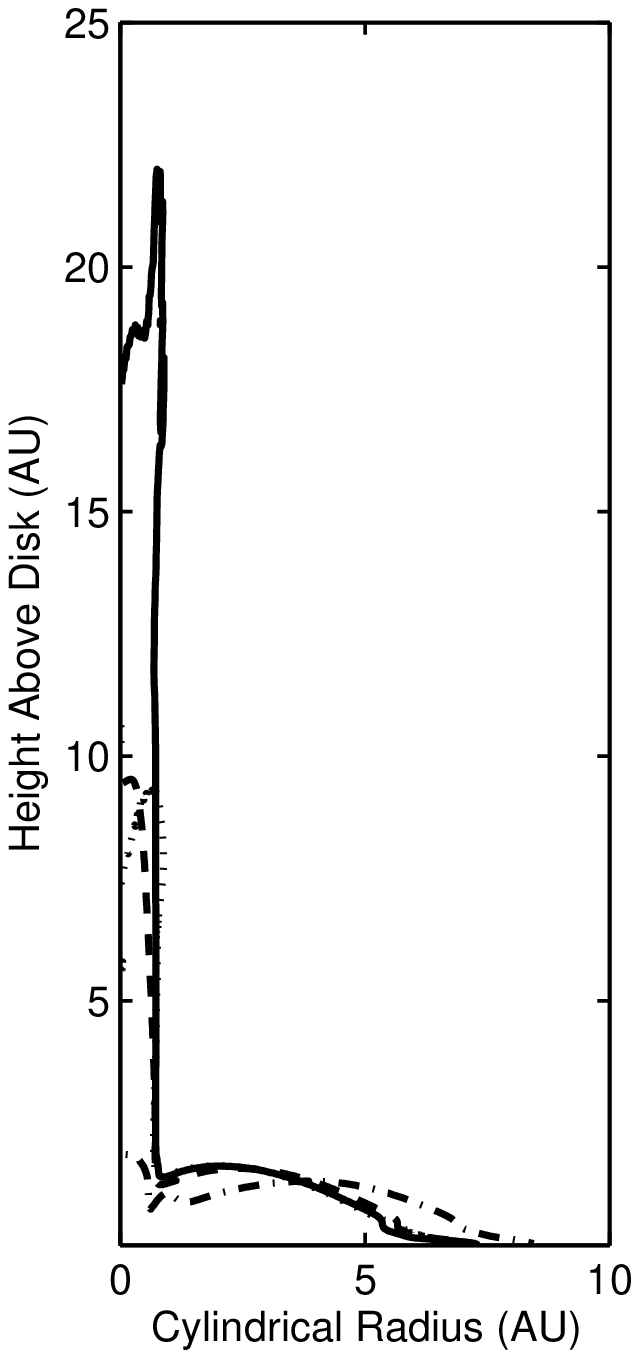}
        \end{center}
        \caption{Plotted are the rescaled density contours (left panel) 
and the Alfv\'en surface (right panel) 
for solutions of different mass loading -- the dotted line
corresponds to $\dot{m}_0 = 0.1$, solid $\dot{m}_0=1$ (reference solution),
dashed $\dot{m}_0=10$, and dash-dotted $\dot{m}_0=100$.  The rough alignment
in both panels demonstrates that the winds are approximately self-similar
in some aspects. }
        \label{fig:f6}
\end{figure*}

\begin{figure*}
        \begin{center}
               \epsscale{1.1}
               \plotone{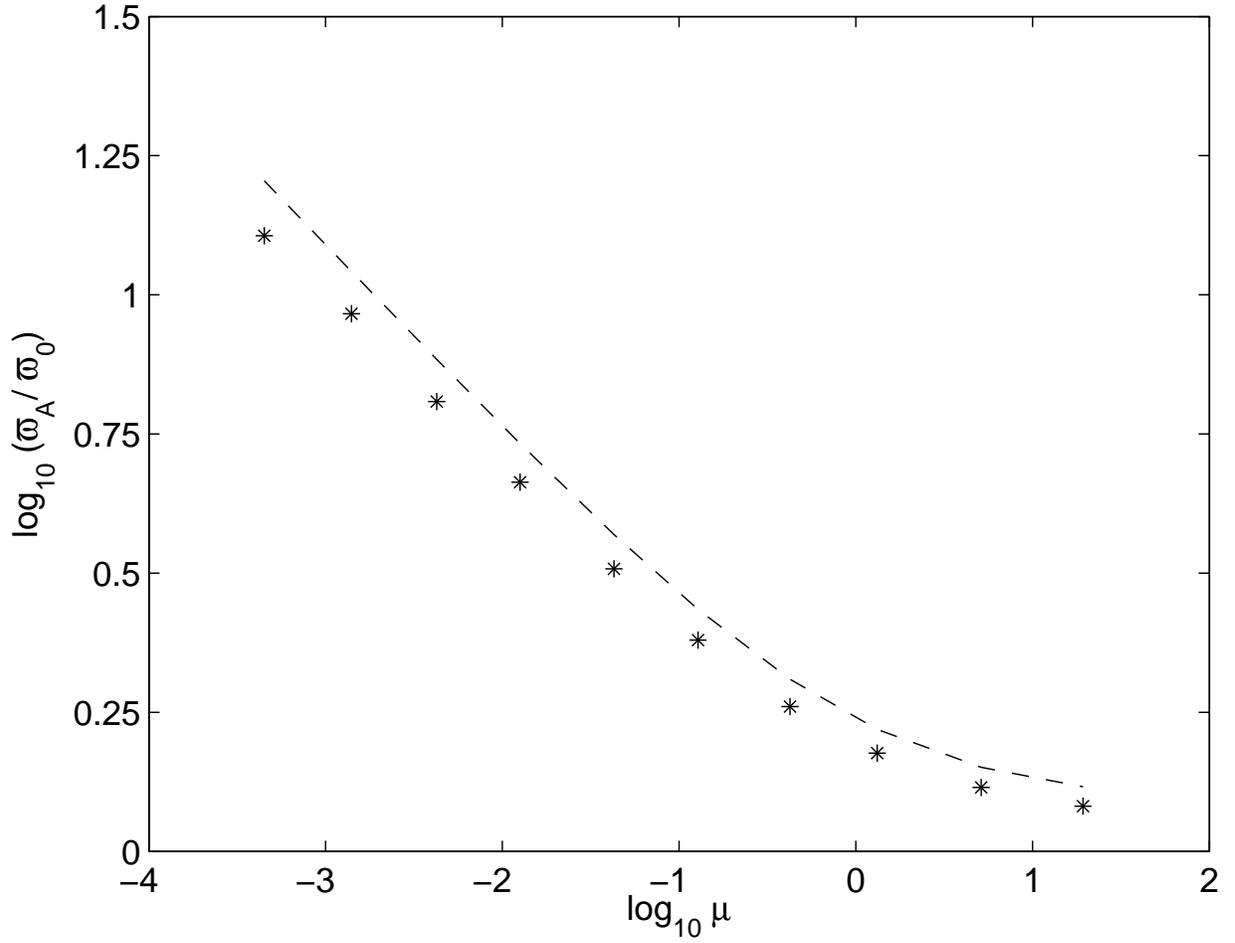}
        \end{center}
        \caption{Log-log plot of the ratio of the cylindrical radius of
          the Alfv\'en point $\varpi_A$ to that of the footpoint 
          $\varpi_0$
          as a function of the mass loading parameter $\mu$ (at the 
          footpoint) for the 50\% streamline.
          The stars are values taken from our simulations while the dashed
          line shows the expected values based on equation\,(71) of 
\citealt{s96}), which was derived for a radial wind geometry.}
        \label{fig:f7}
\end{figure*}

Our wind solutions of different mass loading are approximately self-similar
in several aspects. These include the isodensity contours and the 
Alfv\'en surface, which can be made to line up rather well through 
simple rescaling, as shown in Fig.\,\ref{fig:f6}. The left panel plots
the contour 
of constant number density $5\times 10^4{\dot m}_0^{4/3}\numden$.
The contours of scaled density roughly align. In the right panel,
we plot the Alfv\'en surface with both 
$z$ and $\varpi$ multiplied by ${\dot m}_0^{1/3}$.
Again, most of the surfaces 
line up well, except for the heaviest wind shown. The deviation is 
an indication that the wind may behave differently at the high mass 
loading end.  Interestingly, the
position of the Alfv\'en surface in the simulations closely follows
that predicted for purely radial winds \citep{s96}
\begin{equation}
\frac{\varpi_A}{\varpi_0} = 
\left[\frac{3}{2}\left(1+\mu^{-2/3}\right)\right]^{1/2}
\end{equation}
as shown in Fig.\,\ref{fig:f7}, which includes both ``light'' and 
``heavy'' winds (with $\mu \gsim 1$, to be discussed below). 

\begin{figure*}
        \begin{center}
               \epsscale{1.1}
               \plotone{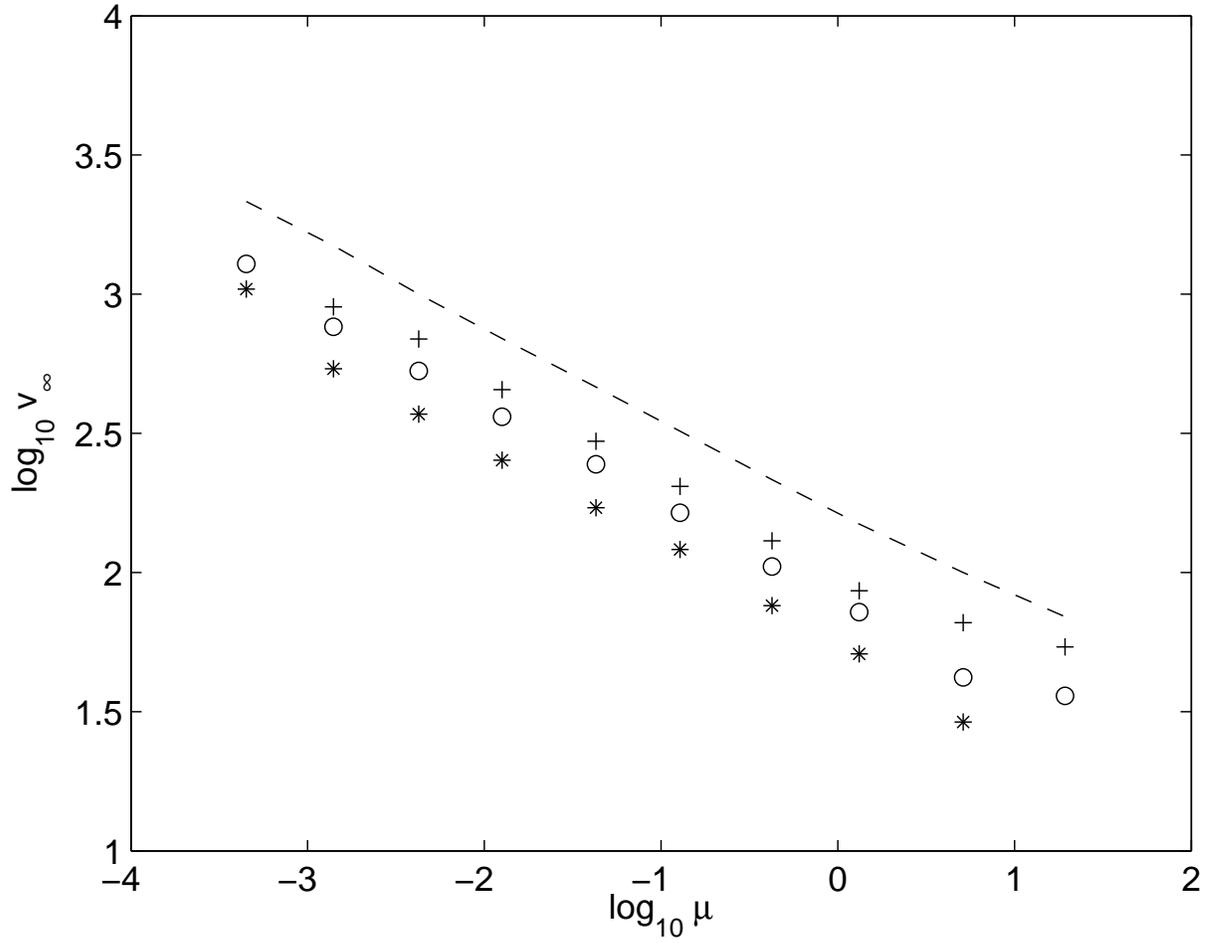}
        \end{center}
        \caption{Log-log plot of the poloidal velocity of the wind along the 
          25 (+), 50 (o), and 75\% (*) streamlines at a radius of $100\AU$
          from the origin as a function of the mass loading. A power-law
          with index $-1/3$ is also displayed for comparison. }
        \label{fig:f8}
\end{figure*}

In Fig.\,\ref{fig:f8},
we plot the poloidal flow speed $v_\infty$ at a spherical radius of 
$10^2\AU$ for three 
representative field lines (enclosing respectively 25, 50, and 75\% 
of the total mass flux) for a number of ${\dot m}_0$. 
The data points follow a power law distribution 
$v_\infty\propto {\dot m}_0^{-\alpha_v}$, with $\alpha_v\approx 1/3$.
At the low ${\dot m}_0$ end, there is a slight 
deviation from power-law; this is 
most likely due to the fact that the fast 
surface approaches the outer edge of the simulation box, which leaves 
little room for the wind to accelerate beyond the fast surface to 
a terminal speed.  There is also a deviation present at large values of
${\dot m}_0$ where the slope of the power appears to become more shallow.  
We note that the index of the power law distribution
of terminal speed with respect to the mass loading is close to 1/3, as
in the simplest case of radial wind geometry (see equation\,[\ref{terminal}]).
The agreement is all the more remarkable considering the fact that the fast
point is no longer at infinity because of wind collimation, and that
there is  substantial increase in flow speed beyond the fast point, by a
factor up to $\sqrt{3}$ for light winds. To be more quantitative, we note 
that at large distances from the launching region, the asymptotic speed 
along any field line can be written in a form  
\begin{equation}
v_\infty = \left({\Omega^2 M_f^2 B_p\varpi^2 \over 
4\pi \kappa }\right)_\infty^{1/3}.
\label{general} 
\end{equation}
The relation is obtained from the definition of the fast Mach number,
the conservation of mass-to-flux ratio (equation\,[\ref{eqn:kappa}]), 
and the flux freezing condition (equation\,[\ref{eqn:omega}]). It is a 
generalization of equation\,(\ref{terminal}) to an arbitrary poloidal
field geometry. 
In the radial field limit, where $M_{f,\infty}=1$ and $B_p\varpi^2$ is
constant along a field line, we have $v_\infty\propto \kappa^{-1/3}
\propto \mu^{-1/3}$, recovering the scaling in equation\,(2). When the flow
collimation is taken into account, neither $M_{f,\infty}=1$ nor  
constant $B_p\varpi^2$ holds along a field line. The fact that the scaling
$v_\infty\propto{\dot m}_0^{-1/3}\propto \mu^{-1/3}$ still holds
approximately implies that the 
combination $(M_f^2 B_p\varpi^2)_\infty$ varies little with mass loading. 
Apparently, the reduction of the (geometric) quantity $B_p\varpi^2$ due 
to the field collimation is more or less offset by the continued 
conversion of magnetic to kinetic energy beyond the fast point, which 
increases the Mach number $M_f$.

\subsubsection{Heavily-Loaded Winds}

\begin{figure*}
        \begin{center}
                \epsscale{0.60}
                \plotone{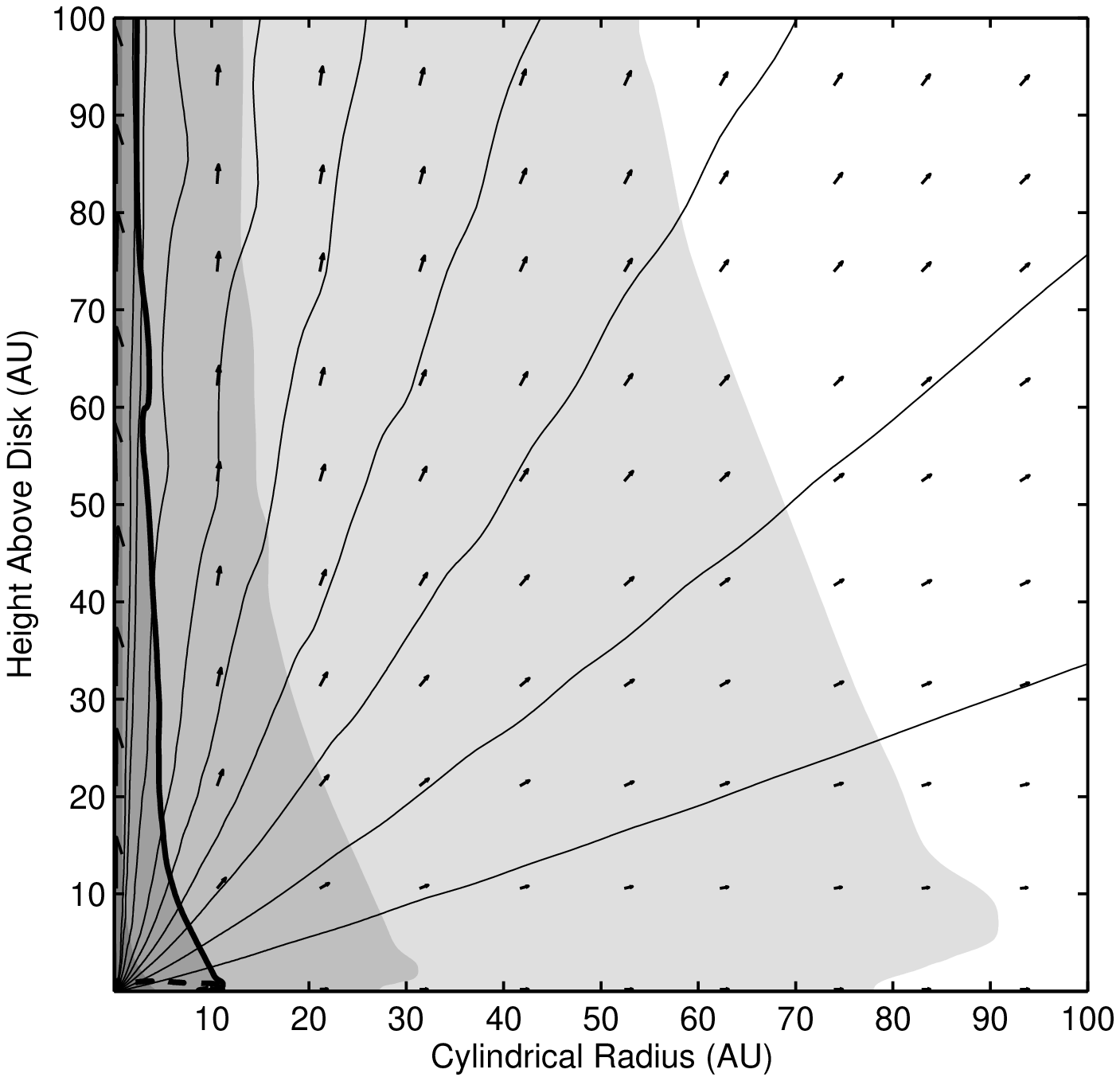}
                \plotone{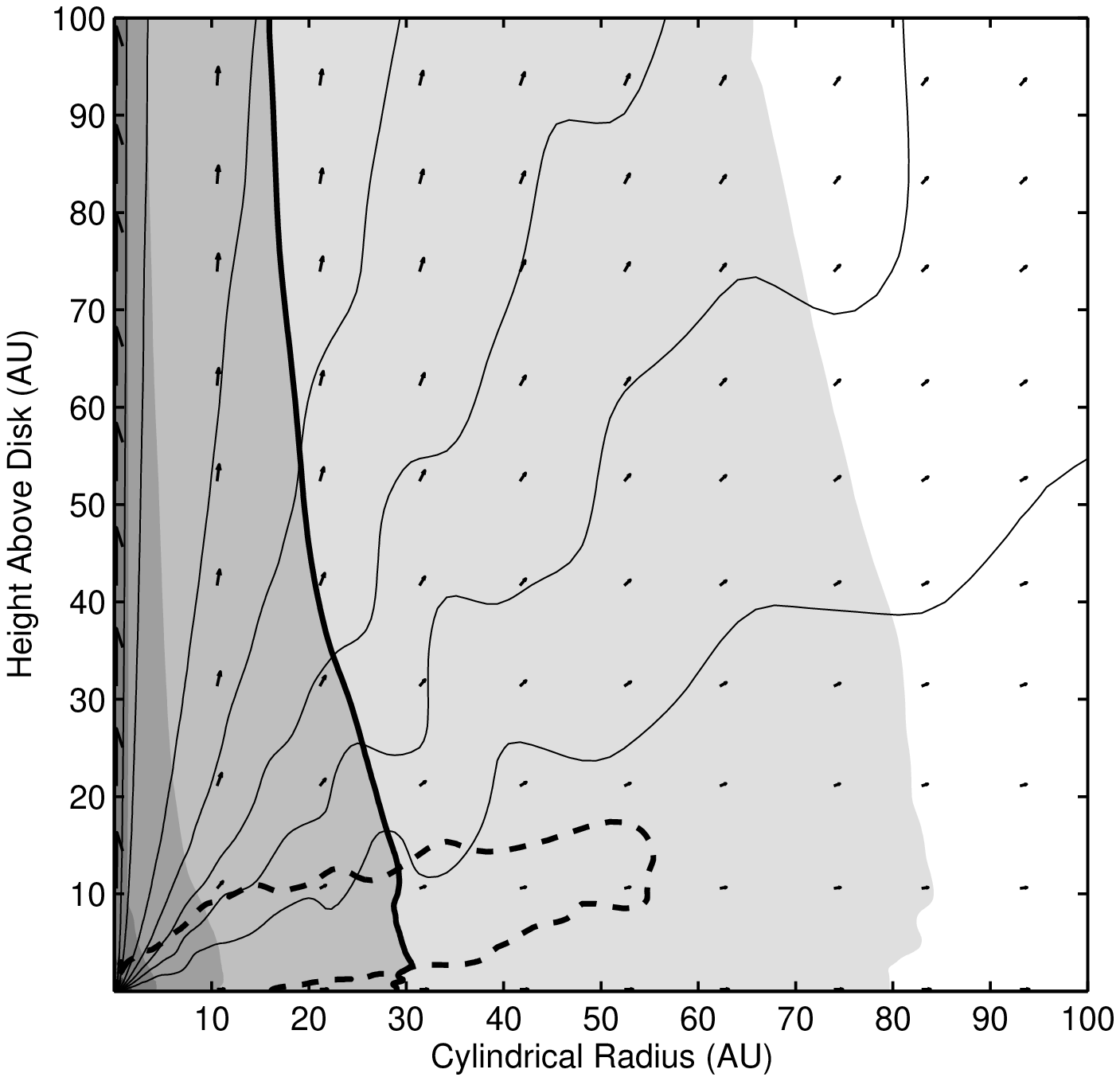}
        \end{center}
        \caption{Wind solutions for the mass loading values of $\dot{m}_0
        =300$ (top panel) and $\dot{m}_0=3000$ (bottom panel), showing 
        the nonsteady flow behavior in the ``heavy'' wind regime. 
        The field lines are shown as thin,
        solid lines, the fast magnetosonic surface as a thick, dashed line,
        and the $10^8\numden$ contour as a thick, solid line.  Greyscale
        indicates number density (one decade per shade), and arrows show 
        the direction and magnitude of the poloidal velocity field. Note
        that the amplitude of oscillation increases with mass loading.}
        \label{fig:f9}
\end{figure*}

Above some mass load, the wind solution remains perpetually unsteady.
The inability to reach steady state begins around ${\dot m}_0=300$, when
ripples of small amplitude start to propagate along some field lines 
from near the launching surface to large distances. As the mass loading
increases, the amplitude of the field line oscillation grows. In the
upper panel of Fig.\,\ref{fig:f9}, we show 
a snapshot of the wind with ${\dot m}_0=300$ 
(corresponding to a characteristic mass loading parameter of $\mu_g=
1.9$), after the wind has settled into a state of finite-amplitude 
oscillation. The origin of the
oscillation is unclear. We suspect that it is related to the large
toroidal magnetic field for a heavily loaded wind, which dominates
the poloidal component from large distances all the way to the base of 
the wind (see discussion in \S\,\ref{heavywind} 
and Fig.\,\ref{fig:f13} below). 
As the mass loading increases further, the amplitude of 
oscillation grows, and the wind starts to turn chaotic. This is illustrated 
in the lower panel of Fig.\,\ref{fig:f9}, 
where the mass loading parameter is set to ${\dot 
m}_0=3000$, and chaotic flow behavior is about to set in. For higher
mass loading, we can trace the chaotic flow behavior to the generation 
of dense blobs near the launching surface, as the poloidal field lines 
bend inwards. The inward bending renders the centrifugal acceleration 
inoperative, as first pointed out in \citet{k00}. In practice, the 
inward bending of the poloidal field lines should also reduce the mass 
loading rate, which is fixed in the current simulation. We will return 
to this point in the discussion section.

We find that the level of nonsteadiness can be reduced by decreasing the 
initial wind injection speed, keeping the mass loading the same. We have
also done a set of simulations using a larger initial injection speed 
($V_o=0.1$ instead of 0.01, see equation\,[\ref{initvel}] for definition
of $V_o$). 
In these simulations the unsteady behavior set in for
lower values of ${\dot m}_0$: oscillations appear at ${\dot m}_0=30$ and for
${\dot m}_0=1000$ the solution was completely chaotic.  Figure\,\ref{fig:f10}
shows a comparison of the $\dot{m}_0=1000$ simulation using the different
injection speeds.  In the top panel we show the $V_o=0.1$ case which is
completely chaotic. Upon reducing $V_o$ to $0.01$ we obtain a relatively
steady
solution (bottom panel) for this mass load. We believe the reduction of 
$v_z$ produces a cleaner result in the cold wind limit. How this result 
would be modified in the presence of a dynamically important thermal 
pressure near the base of the wind is a subject of future investigation. 

\subsection{Variation in the Distribution of Mass Loading}
\label{loaddist}

\begin{figure*}
        \begin{center}
               \epsscale{1.1}
               \plotone{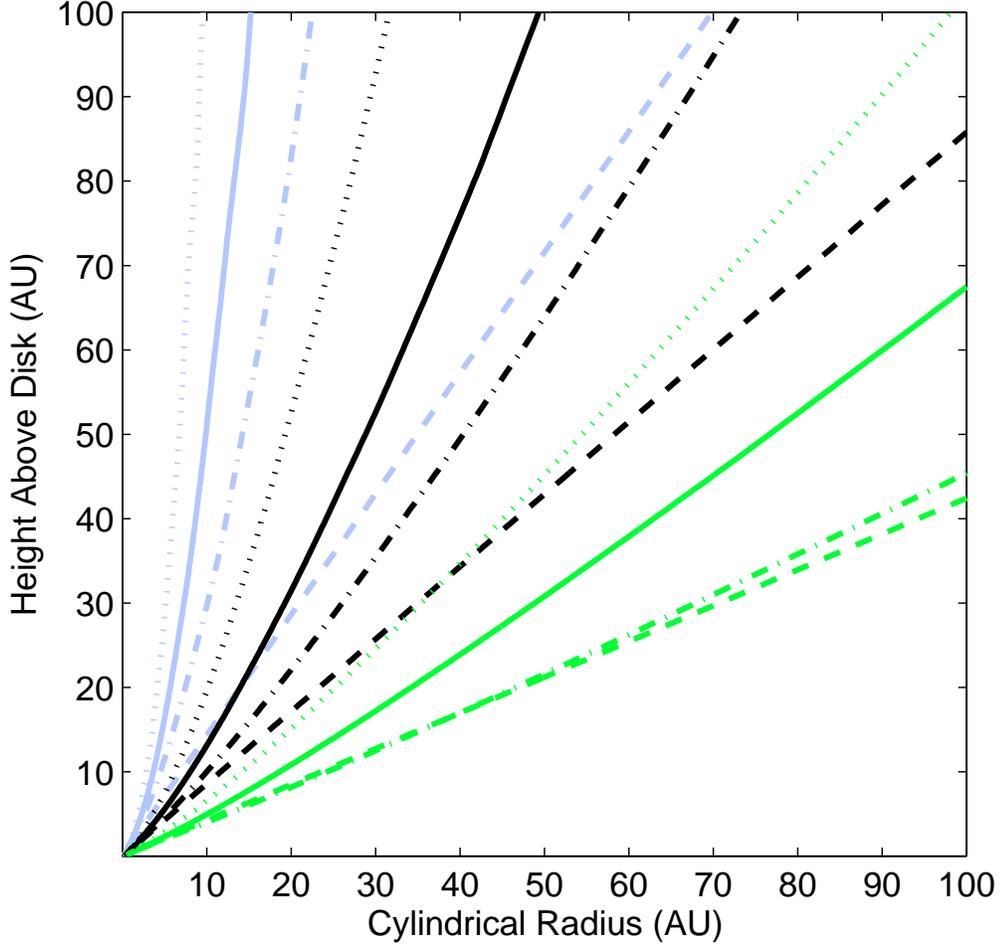}
        \end{center}
        \caption{Three field lines plotted for simulations with
same total mass flux, but differing mass loading distributions
 (25\% line - blue, 50\% line - black, 75\% line - green).
The dotted line corresponds to $\alpha_m = 1$, solid $\alpha_m=2$
(reference solution), and dash-dotted $\alpha_m=3$.  The thick dashed line
shows the initial field for all three cases.  The degree of magnetic field
collimation clearly increases with decreasing $\alpha_m$.  For the steep
$\alpha_m=3$ mass loading, the mass load at the outer edge of the launching
surface is so slight that it has hardly bent the field away from its initial
configuration.}
        \label{fig:f11}
\end{figure*}

\begin{figure*}
        \begin{center}
               \epsscale{1.1}
               \plotone{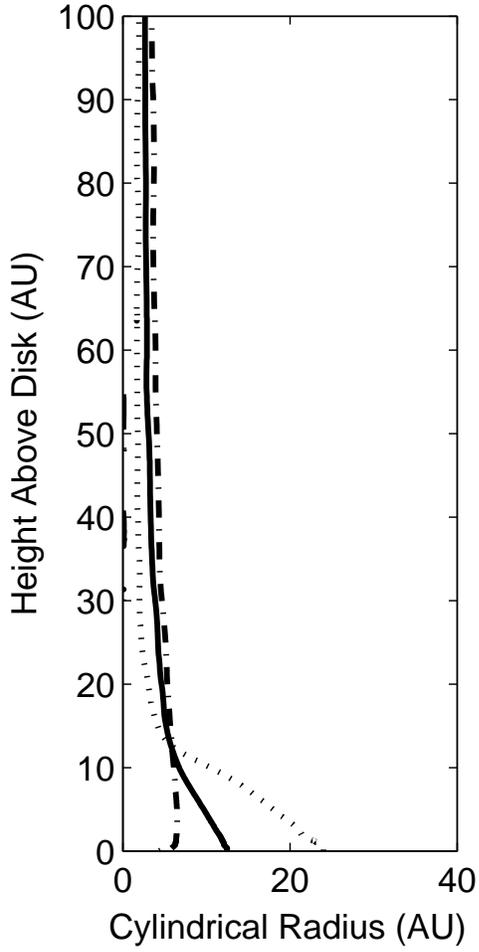}
        \end{center}
        \caption{Density contours ($n_H=5\times 10^4\numden$) for different
        mass loading distributions.  The solid line corresponds to $\alpha_m
        =2.0$ (reference solution), the dotted line $\alpha_m=1.0$, and 
        dash-dotted line $\alpha_m=3.0$.  The solutions have all been scaled
        such that the total mass flux is $10^{-8}\solarmassyear$.}
        \label{fig:f12}
\end{figure*}

The reference solution and its variants discussed in the last two 
subsections all have a dimensionless mass loading parameter $\mu$ 
more or less uniform across the surface of magnetocentrifugal wind 
launching. There is the possibility that the conclusions drawn may 
depend on this special feature of the solutions. To investigate 
whether this is the case, we have carried out several simulations 
in which $\mu$ varies substantially on the launching surface. The
desired variation can in principle be obtained by changing either
the magnetic flux distribution or the distribution of mass loading.
We choose the latter, by varying the exponent $\alpha_m$ for the 
density distribution on the launching surface from 0.5 to 3, but 
fixing the total mass flux from each side of the disk to $\dot{M}_w
=10^{-8}\times(B_0/19.2\gauss)^2\solarmassyear$. The results are 
shown in Figs.\,\ref{fig:f11} 
and \ref{fig:f12}. Fig.\,\ref{fig:f11} displays the steady state 
field configurations for three of the simulations. For each steady
state solution, we plotted three field lines from the same three
footpoints on the launching surface for all cases. The degree of 
collimation of field lines from a given footpoint is anticorrelated
with the steepness of the mass loading slope, and, in fact, for the
steepest simulation ($\alpha_m=3$) the last field line plotted is 
only barely collimated compared to the potential field configuration. 
The weak collimation is a result of small mass loading between that 
field line and the equatorial plane. 

Density contours for the three simulations are shown in Fig.\,\ref{fig:f12}.
Although the steepest mass loading simulation has the least collimated 
field from a given footpoint, it does possess the most ``jetlike'' 
density contours (i.e.\ the least amount of bulging at the base of 
the wind) --- undoubtably due to most of the plasma being centrally 
concentrated on the launching surface. This solution is more attractive
than the other two in explaining the nearly cylindrical appearance 
observed in some YSO jets, such as HH 30 \citep{bur96}. 

We have carried out a series of simulations for $\alpha_m=1$ and $\alpha_m
=3$ where we have varied the total mass flux, keeping the distribution 
(i.e., $\alpha_m$) the same, as in \S\,\ref{loadmagn}. We find that,
as the mass
loading increases, the shallower mass distribution simulations with 
$\alpha_m=1$ become unsteady sooner. The lower maximum load for steady
winds in this case is probably a reflection of the fact that at large disk 
radii the rotation is slower and the field strength is smaller, both 
of which make the acceleration of a heavy load magnetocentrifugally
difficult. Nevertheless, we have been able to verify that variations 
in mass loading still produce the same power law scalings in density 
contour and Alfv\'en surface locus for these alternate mass distributions 
as for the reference simulations. This indicates our results are not 
simply a fortuitous choice of mass distribution. Similarly the terminal 
velocities and magnetic lever arms follow the expected relations (see 
equations\,[\ref{mu_g1}] and [\ref{mu_g2}]).

\section{DISCUSSION AND CONCLUSIONS}
\label{conclusion}

\subsection{Mass Load Limit and Heavy Magnetocentrifugal Winds}
\label{heavywind}

\begin{figure*}
        \begin{center}
                \epsscale{0.80}
                \plotone{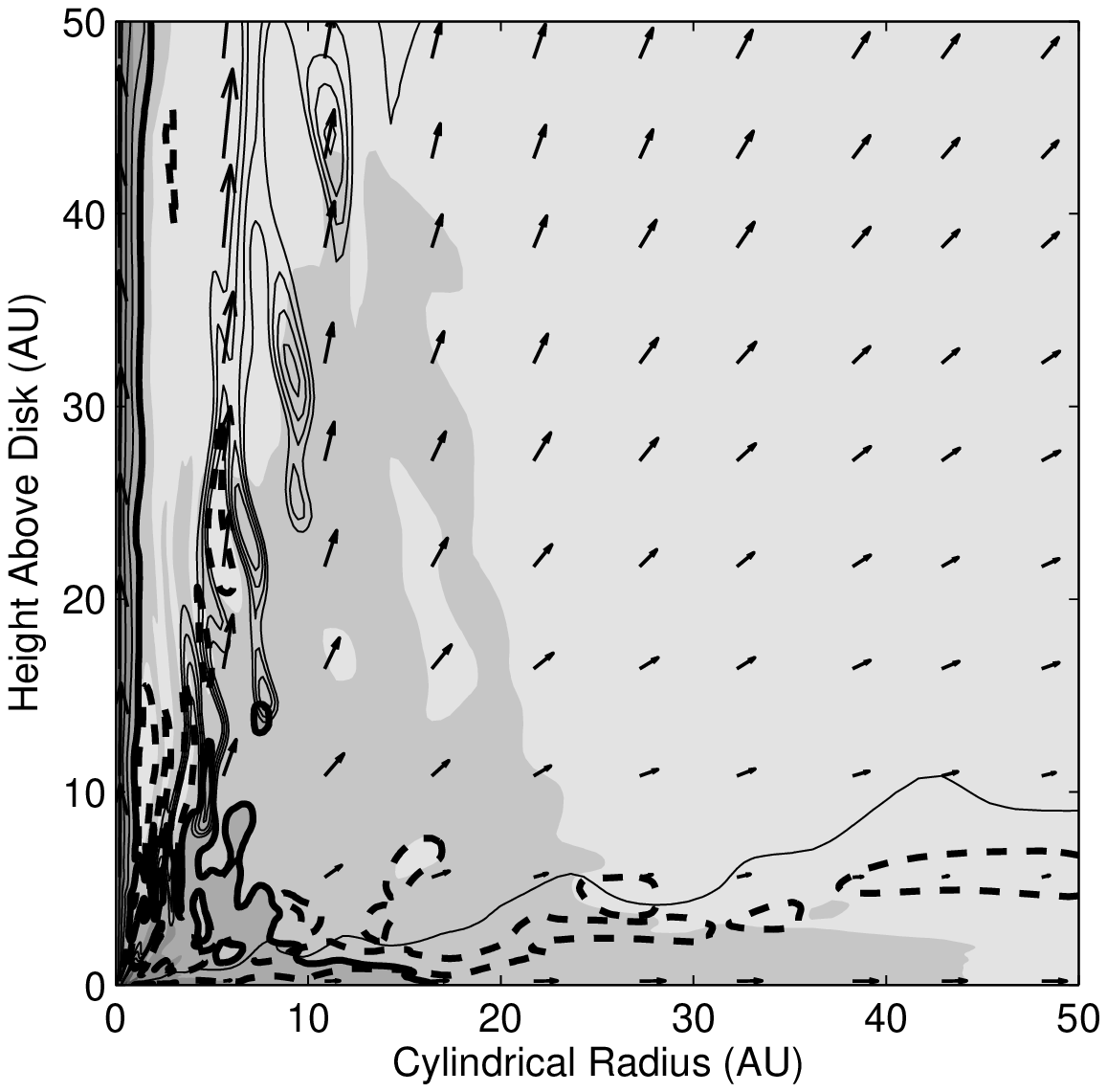}
                \plotone{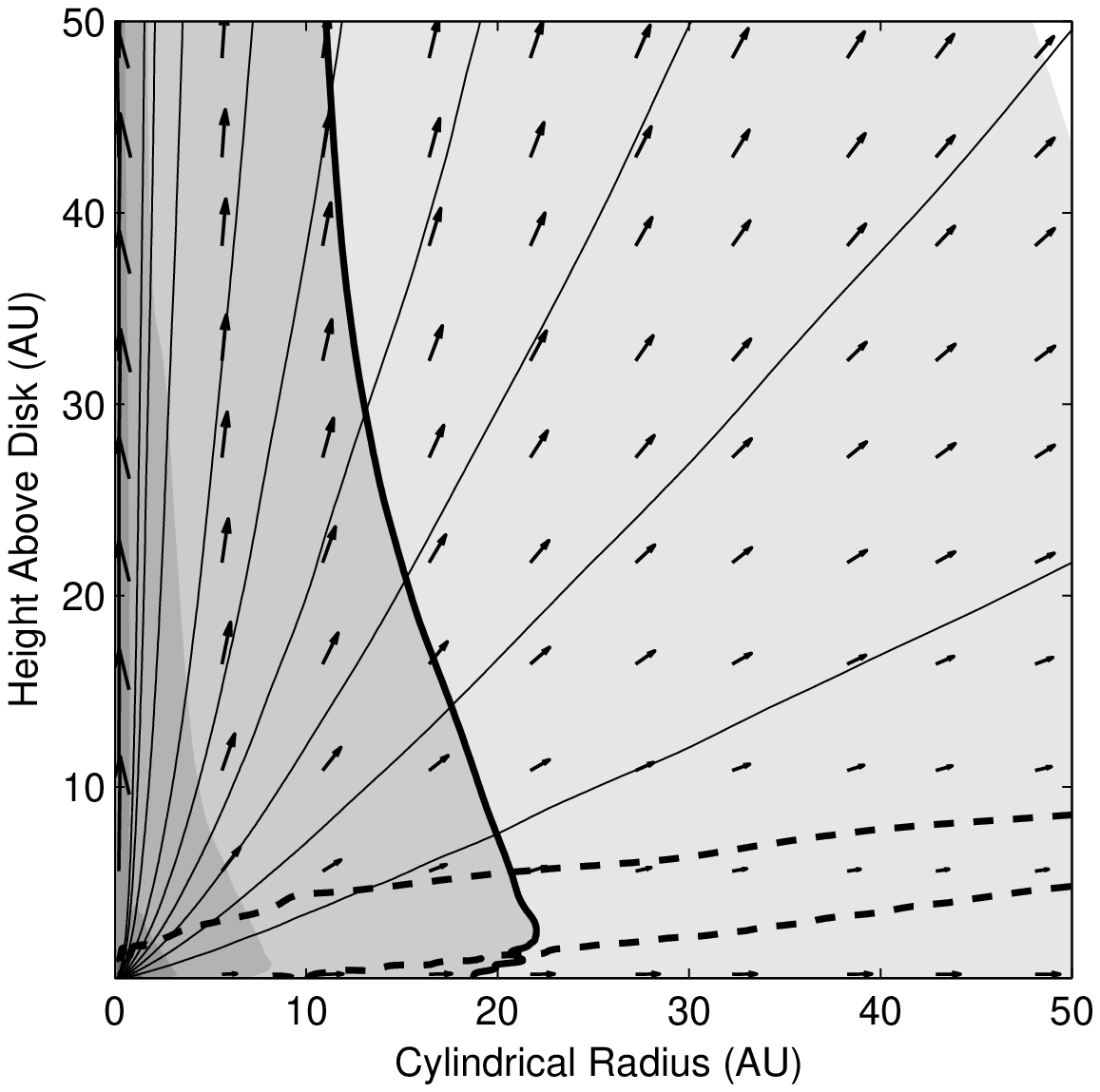}
        \end{center}
        \caption{Wind solutions for mass loading value $\dot{m}_0=1000$ with
        initial injection speed $V_o=0.1$ (top panel) and $V_o=0.01$ (bottom
        panel). The field
        lines are shown as thin,
        solid lines, the fast magnetosonic surface as a thick, dashed line,
        and the $10^8\numden$ contour as a thick, solid line.  Greyscale
        indicates number density (one decade per shade), and arrows show the
        direction and magnitude of the poloidal velocity field.}
        \label{fig:f10}
\end{figure*}

A novel finding of our paper is that, for a given wind launching magnetic 
field, there exists a maximum mass load beyond which cold, steady state 
solution does not exist. For a relatively large wind injection speed of 
$10\%$ 
Keplerian, the maximum corresponds to a dimensionless parameter $\mu
\sim 1$, which is roughly where the transition from ``light'' to ``heavy'' 
wind occurs. With the reduction of injection speed to $1\%$ Keplerian we can
obtain rather steady solutions up to $\mu \sim 10$. The tendency for the 
solutions in the heavy wind regime to 
remain unsteady may not be too surprising, given that their magnetic fields
are dominated by the toroidal component all the way to the launching 
surface. This is illustrated in Fig.\,\ref{fig:f13} for the solution 
shown in lower 
panel of Fig.\,\ref{fig:f10}, 
where the mass loading is 1000 times higher than the 
reference solution. The toroidal field in the heavy wind
is everywhere greater than the poloidal field in the 
magnetocentrifugal part of the wind.  For comparison, the toroidal field
in the light, reference solution starts to exceed the poloidal field
only well beyond the launching surface.   At the launching surface itself, the ``heavy''
wind has a ratio $|B_\phi|/B_p$ of 4, compared to a value of $0.2$ for the 
``light'' wind. The toroidally dominated magnetic 
configurations are prone to pinch instability in lab experiments 
\citep{ks54}. A possible indication of the onset 
of this type of instability is the waviness of field lines in solutions with
mass loads near but below the threshold for transition to chaotic
flows (see Fig.\,\ref{fig:f9}). The fact that the threshold increases with 
decreasing injection speed at the launching surface points to another 
possibility for the nonsteadiness that involves the gravity of the central
object --- loss of mechanical balance near the launching surface for 
heavy winds. 

The reason for the loss of balance can be seen from equation
(\ref{eqn:omega}) for the
conservation of angular velocity $\Omega$ along any field line in steady
state. The equation can be cast into 
\begin{equation}
v_\phi=\Omega \varpi + v_p B_\phi/B_p.
\label{rotation}
\end{equation}
At the launching surface, the first term on the right hand side is set to 
the Keplerian speed in our simulations, and the second represents 
the speed that the wind fluid lags behind the Keplerian rotation (note 
that $B_\phi$ is negative). For a given injection speed $v_p$, the lag 
increases with the twisting of the field lines (i.e., the ratio of $\vert 
B_\phi\vert/B_p$), which in turn increases with mass loading. It is 
conceivable that beyond some threshold in mass load, the fluid rotation 
near the launching surface becomes too sub-Keplerian for the centrifugal 
force to balance the inward pull of the central gravity \citep{k00}. 
The force 
imbalance would lead to a radial infall of wind material, which is indeed 
observed in the initial phase of the development of chaotic flows, such 
as the one shown in upper panel of Fig.\,\ref{fig:f10}. The development 
of the chaotic behavior is reminiscent of that of the channel flows 
observed in simulations of weakly magnetized accretion disks 
(J. Hawley, priv.\ comm.). As one decreases the injection speed, the 
lag speed is decreased for a given magnetic field twist. This is 
consistent with our finding that decreasing the injection speed tends 
to make a heavy wind more steady. By choosing a small enough injection 
speed we were able to obtain more or less steady solutions well into the 
heavy wind regime, with $\mu \gg 1$. The fact that such heavy wind 
solutions can be obtained from {\it time-dependent} simulations suggests 
that they are stable to, or at least not disrupted by, axisymmetric (pinch) 
instabilities, despite their magnetic fields being completely dominated 
by the toroidal component. Whether they can resist disruption by 
non-axisymmetric (kink) instabilities in 3D remain to be determined. 

Even for an arbitrarily small injection speed, we expect the wind solution 
to become chaotic beyond a certain mass load. The reason is still 
related to equation (\ref{eqn:omega}), which applies to both the launching 
surface and above. For a slowly injected wind, acceleration above the 
launching surface increases the poloidal speed, causing the fluid to rotate 
significantly below the rate needed for radial force balance when
the magnetic field becomes toroidally dominated. Material 
fed into the wind starts to fall toward the central star, dragging 
magnetic field lines along with it. The inward bending of field lines 
further reduces the centrifugal acceleration and thus exacerbates the 
force imbalance, again reminiscent of the development of channel flows
in weakly magnetized disks through magneto-rotational instability 
\citep{hb91}. This positive feedback can be cut off if 
the mass loading (which is fixed in our simulations) is allowed to vary 
in response to the field bending. The mass loading is expected to drop 
drastically when the (poloidal) field lines bend within roughly $30\degr$ 
of the disk normal, the minimum angle for mass-loading by the 
magnetocentrifugal mechanism (\citealt{bp82}; \citealt{ogil01}). 
The reduction is a natural way for the wind to self-limit its amount 
of mass loading. Quantifying this process requires a detailed treatment 
of the disk-wind coupling, which is beyond the scope of the present
work (see, e.g., the early analytic work of \citealt{wk93} and recent 
numerical simulations of \citealt{ck02}). 

\citet{op99} addressed the issue of mass loading through numerical 
simulations. They found a transition from steady to non-steady wind 
solution as the mass load is decreased below some critical value; 
this trend is the opposite of the one we find. One possibility 
for the difference lies in the simulation setup at the launching 
surface. In their simulations, the toroidal field strength 
is prescribed at the disk surface, along with the magnetic flux 
distribution. This prescription essentially fixed the amount of 
energy (and angular momentum) flux extracted by the rotating 
magnetic fields from the disk proper (see the last term in the 
expression, equation~[10], for the specific energy $E$). In general, 
these prescribed amounts cannot be carried away from the disk by 
the loaded wind material, as illustrated in \citet{op97a}. They 
found a discontinuity in $B_\phi$ between the disk and 
the active zones in a test simulation with $B_\phi$ set to zero on 
the disk. The discontinuity indicates to us that the toroidal 
field near the base of the wind is trying to respond to stresses 
further out, while being constrained by a (formally) mismatched 
boundary condition. In our simulations, the toroidal field is 
allowed to adjust in response to the stresses in the wind both on 
the launching surface 
and in the active zones. We are able to obtain steady solutions 
for very low mass loads, limited only by the size of the simulation 
box (since the fast surface, which needs to be enclosed within 
the box to minimize boundary effects, increases in size as mass 
load decreases), and by computation time, because a lighter wind 
has a larger Alfv\'en speed, which requires a smaller timestep to 
simulate. Another possibility for the difference is that our 
magnetocentrifugal wind becomes super fast-magnetosonic within 
the computational box, whereas in their simulations a significant 
fraction of outflow remains sub-fast up to the outer boundary. 
One worry is that, in such sub-fast regions, signals can propagate 
from the outer boundary to the launching surface. In the numerical
experiments of KLB99, the coupling between the outer boundary and
launching surface appears capable of driving the magnetocentrifugal
wind unsteady or even chaotic.

The non-steadiness in our high mass load winds has a different 
origin: it occurs when the field is in some sense too weak to 
accelerate the prescribed mass load steadily. In our 
interpretation, the maximum in mass loading has its root in the 
inability of a heavily loaded wind to find a stable cross-field force 
balance. Such a force-balance cannot be accounted for in 1D wind models 
along a prescribed flux tube. Even 2D (axisymmetric) models obtained
by directly solving the steady wind equations, such as those of 
\citet{sa87} and \citet{ns94}, fail to uncover this maximum. This is 
probably because the construction of steady wind solutions directly 
is difficult, and it has not been possible to explore a large parameter 
space as we did in this paper using the time-dependent simulations. 
One type of steady solutions for which 
parameter exploration is feasible is the self-similar solutions of 
\citet{bp82}. However, the cross-field force balance of such solutions 
must be interpreted with care, because of singularities near the axis 
and at the infinity. For example, \citet{sa87} pointed out that the 
``standard'' solution of Blandford \& Payne is less collimated than 
the potential magnetic field with the same flux distribution on the 
disk, as the result of a singular current density on the axis. Also, 
the self-similar solutions tend to recollimate at large distances, 
which is not observed in our non-self-similar steady solutions. 

\subsection{Scaling Laws for Magnetocentrifugal Winds}

\begin{figure*}
        \begin{center}
                \epsscale{0.80}
                \plotone{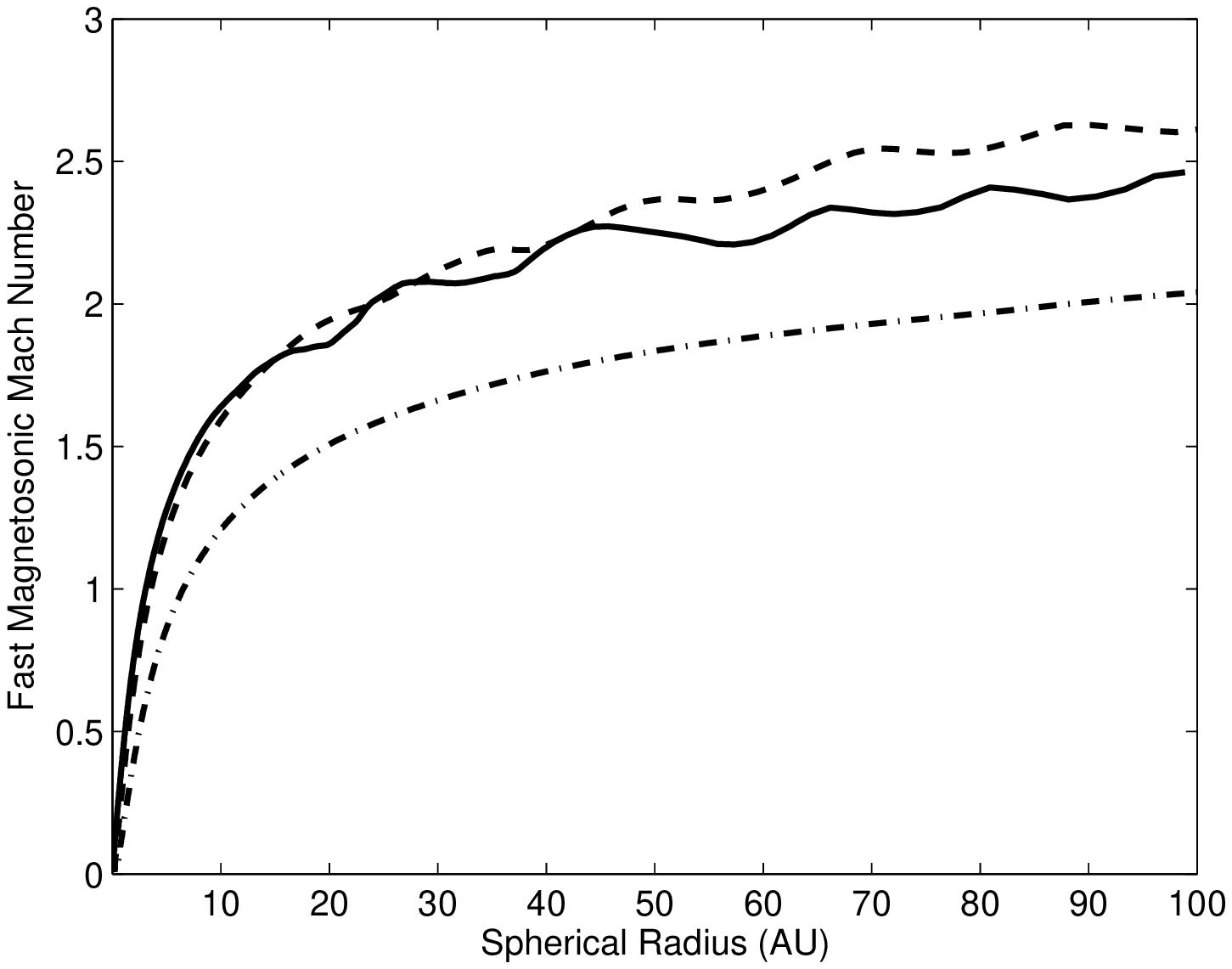}
                \plotone{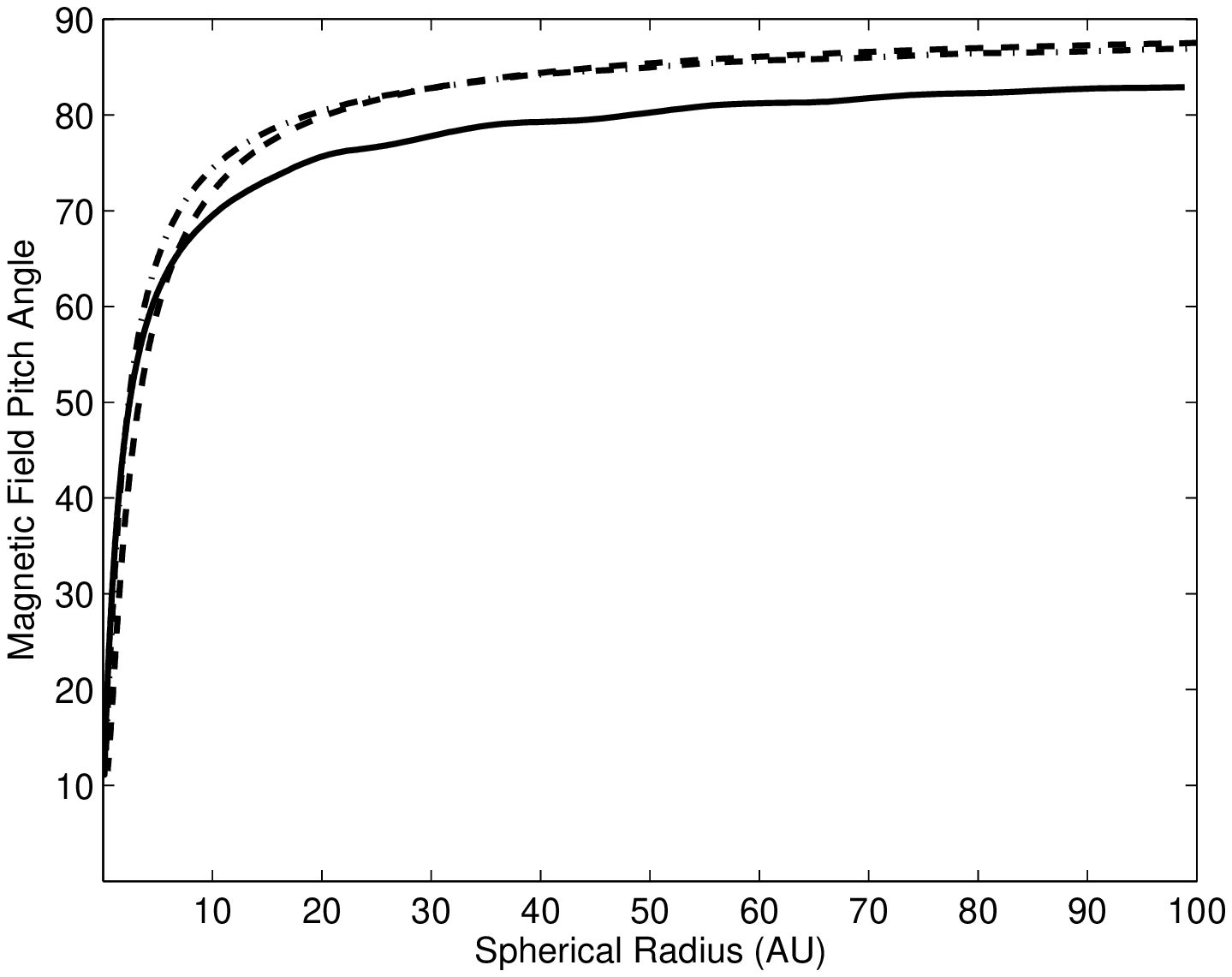}
        \end{center}
        \caption{Reference simulation, showing the fast magnetosonic
Mach number ${\cal M}_{f}$ (top panel) and the pitch angle
$\theta=\tan ^{-1}(|B_\phi|/B_p)$ of magnetic field (bottom panel)
as a function of spherical radius $r$ along three representative 
magnetic field lines. The solid line corresponds to the 25\% mass flux 
enclosed field line, the dashed line the 50\% line, and the dash-dot 
line the 75\% line.  }
        \label{fig:f4}
\end{figure*}

Self-collimation of streamlines is a basic feature of magnetocentrifugal 
winds. The general expectation is that, for a given distribution of 
magnetic flux, the wind becomes better collimated as the amount of mass 
loading increases (e.g., \citealt{pp92}). This expectation 
is borne out by our solutions (see Fig.\,\ref{fig:f4}). The increase 
in the degree of collimation with mass load appears to be slow; fitting 
a power-law distribution to the cylindrical radius of a representative 
streamline (at a height of $100\AU$) listed in Table\,\ref{tab:t1} as 
a function of mass load yields a rough relationship, $\varpi_j\propto 
{\dot M}_w^{0.1}$. This relationship does not apply to all streamlines. 
In particular, the first (on axis) and last (on equator) streamlines 
are the same for all of our winds (by design). The space-filling 
constraint tends to discourage rapid streamline collimation, 
particularly for winds launched from a region that is small compared
to the computational domain. 
Nevertheless, it is difficult to predict the location of each individual 
streamline a priori. This needs to be determined through numerical 
computation. 

A somewhat surprising result is that the wind speed along any given 
streamline scales with the mass load as ${\dot M}_w^{-1/3}$, 
originally 
derived for radial geometry, despite the fact that the streamline 
collimates to different degrees for different mass loads. A related 
result is that the Alfv\'en radius follows closely the analytical 
relation $\varpi_A (\mu)$ for a radial wind. Both results indicate 
that the energy and angular momentum extraction from the disk and the 
wind acceleration are insensitive to the relatively small differences 
in the degree of flow collimation for various mass loads. The fact 
that the appropriately scaled density contours and Alfv\'en surfaces 
closely align implies that magnetocentrifugal winds of different mass 
loads are approximately self-similar. 

The distribution of mass loading also affects flow collimation. We have
demonstrated that loading more material in the outer part of the wind
tends to produce a better collimated streamline from a given location
on the disk. This is probably due to a preferential increase in the 
toroidal field strength of the outer part, which enhances the cross-field 
gradient of the product $B_\phi\varpi$, which is largely responsible 
for flow collimation. On the
other hand, concentrating more material near the inner edge of the
Keplerian disk tends to produce better collimated density contours. 
The fact that jet-like density structure can be produced naturally, 
together with the detection of rotation signatures in T Tauri jets 
(\citealt{bac02}; \citealt{coff04}), lends strong support 
to the magnetocentrifugal model of jet formation (\citealt{shu95}; 
\citealt{and03}). 

To summarize, we have numerically obtained axisymmetric magnetocentrifugal 
wind solutions for a wide range of mass loading for a given wind-launching 
field configuration. The range is limited from below by computation 
time, and from above by a form of instability that causes the wind 
to become chaotic. It testifies to the robustness of the 
magnetocentrifugal mechanism for outflow production. Whether the 
mass loading in actual astrophysical systems, such as YSO winds, 
covers as wide a range remains to be determined. It will probably 
be limited by the detailed physics of mass supply from the disk, 
and by the ability of the wind to withstand disruption by instabilities 
in 3D. 

\acknowledgements
This work was supported in part by NASA grants NAG 5-7007, 5-9180, 
and 5-12102, NSF grant AST 00-93091, and the
F.H. Levinson Fund of the Peninsula Community Foundation.  For the
production runs of our simulations, we utilized the Xeon Linux
Supercluster of the National Computational Science Alliance, which 
is supported under grant MCA03S038. We thank the referee for comments 
that improved the presentation of the paper.

\begin{deluxetable}{ c c c c c } \tabletypesize{\small}
        \tablewidth{450pt.}
        \tablecaption{Wind Parameters as a Function of Mass Load 
          $\dot{M}_{w}$}
        \tablehead{ \colhead{$\dot{M}_{w}$ ($10^{-8}$ M$_{\sun}$ yr$^{-1}$)}
           &\colhead{$\mu_g$} & \colhead{$v_{\infty}^a$ (km s$^{-1}$)} 
          & \colhead{$M_{f,\infty}^a$}
          &\colhead{$\varpi_{j}^a$ (AU)}}
        \startdata
        3000.0 & 19.0     &  36    &  2.26  &  14             \\
        1000.0 & 6.3           &  42    &  2.43  &  19        \\
        300.0  & 1.9      &  72    &  2.83  &  26             \\
        100.0  & 0.63   &  105   &  2.78  &  30               \\
        30.0   & 0.19    &  164   &  2.92  &  30              \\
        10.0   & 6.3$\times 10^{-2}$ &  245   &  2.83  &  30   \\
        3.0    & 1.9$\times 10^{-2}$ &  363   &  2.70  &  32  \\
        1.0    & 6.3$\times 10^{-3}$ &  530   &  2.62  &  32  \\
        0.3    & 1.9$\times 10^{-3}$ &  763   &  2.47  &  36 \\
        0.1    & 6.3$\times 10^{-4}$ &  1285  &  2.28  &  48  \\
\enddata
        \label{tab:t1}
\tablenotetext{a}{Evaluated on the 50\% mass flux enclosing field line.}
%\tablecomments{etc...}
\end{deluxetable}

\clearpage

\clearpage

\clearpage

\clearpage

\clearpage

\clearpage

\clearpage

\clearpage

\clearpage

\clearpage

\clearpage

\clearpage

\clearpage

\begin{figure*}
        \begin{center}
                \epsscale{0.7}
                \plotone{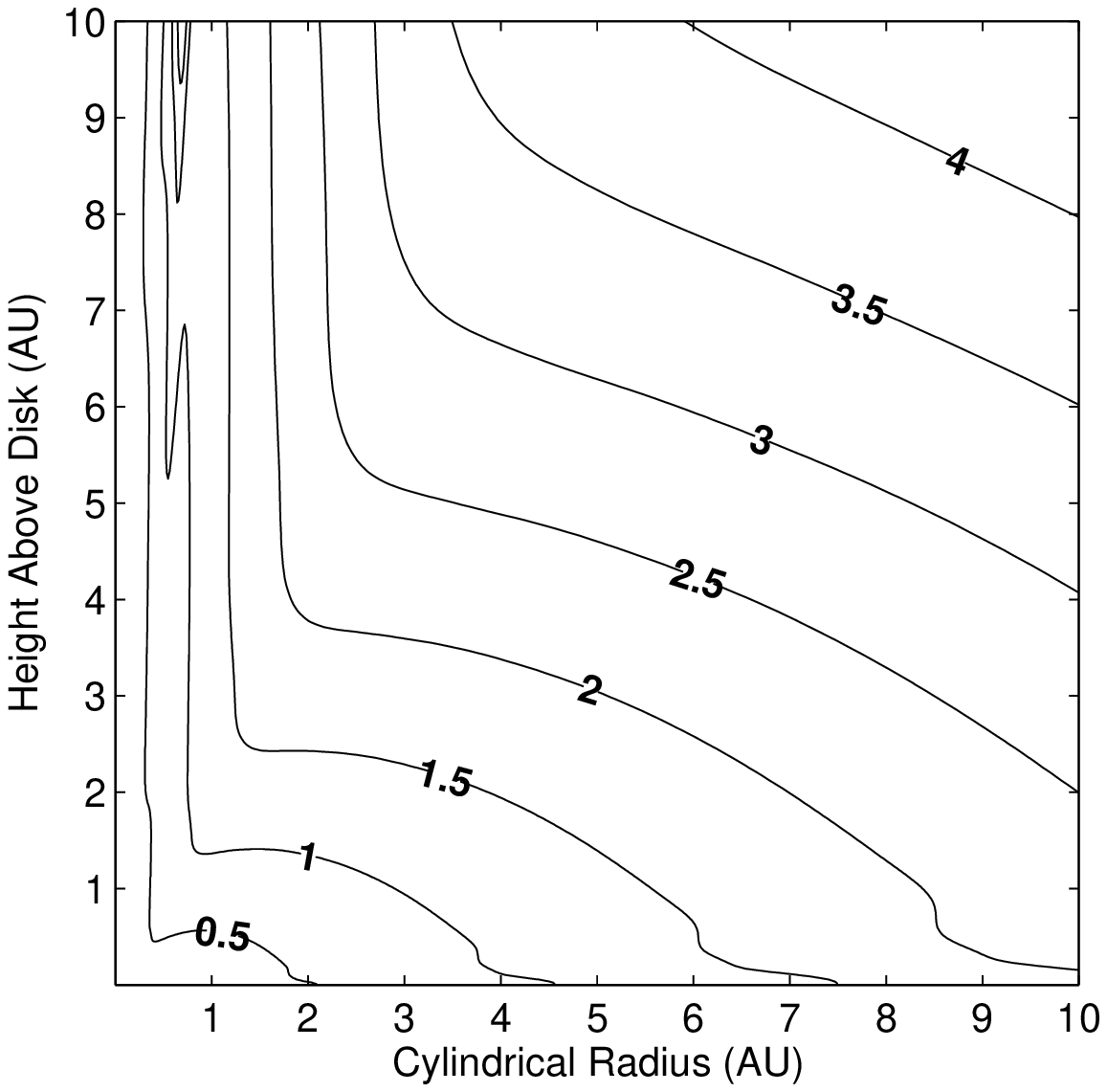}
                \plotone{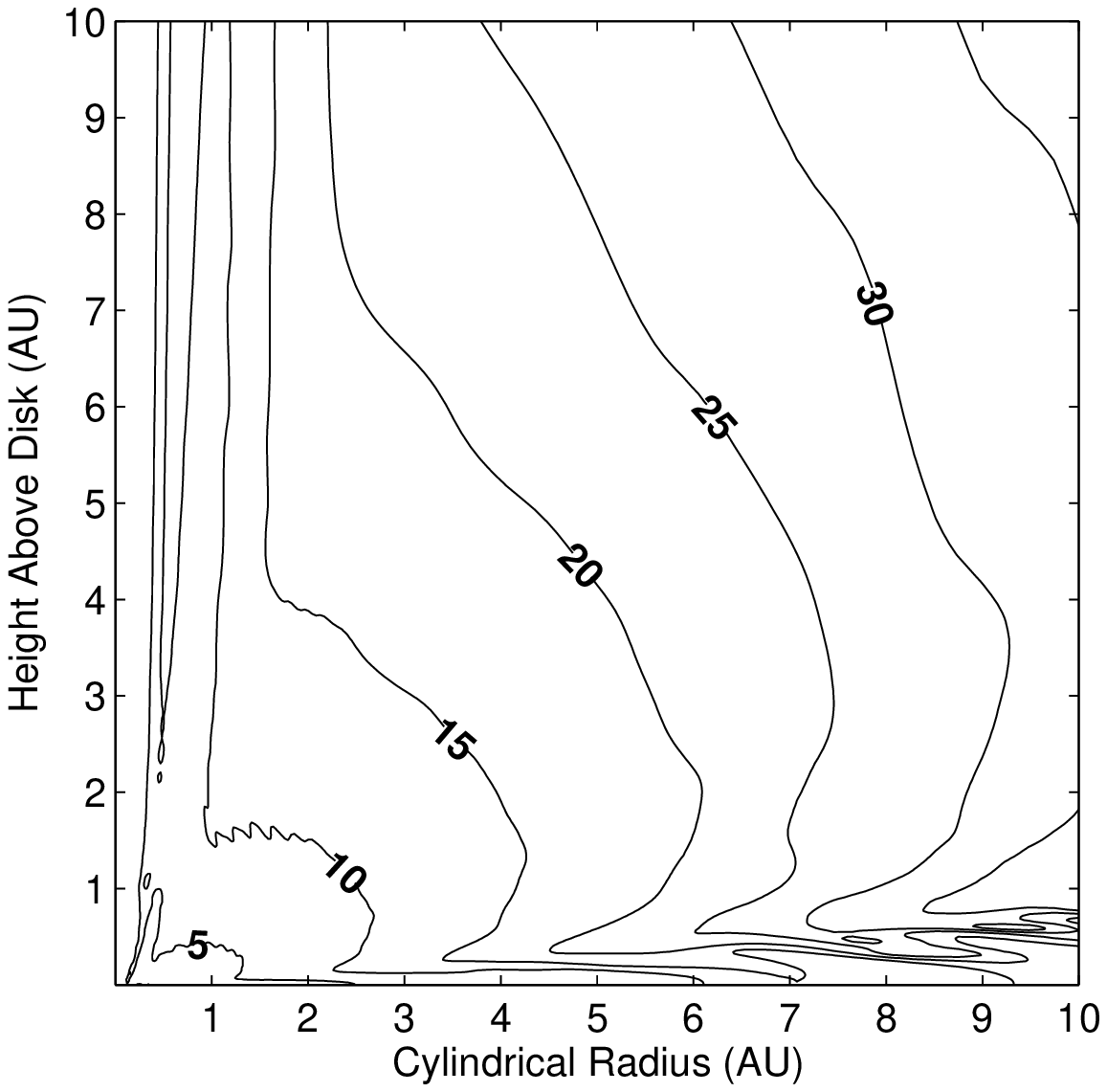}
        \end{center}
        \caption{Contours of the ratio $|B_\phi|/B_p$ for the inner
        regions of the reference
        simulation with $\dot{m}_0=1$ (top panel) and the heavily loaded, 
        $\dot{m}_0=1000$, 
        simulation (bottom panel).  Whereas the reference wind is initially
        dominated by the poloidal magnetic field, the heavily loaded 
        wind is clearly dominated by the toroidal field all the way 
        to the launching surface.}
        \label{fig:f13}
\end{figure*}

\clearpage

\end{document}